\documentclass[reprint,aps,prd,superscriptaddress,amsmath,amssymb,nofootinbib,floatfix]{revtex4-2}

\usepackage{graphicx}
\usepackage{dcolumn}
\usepackage{bm}
\usepackage{hyperref}
\usepackage{subcaption}
\usepackage{booktabs}

\begin{document}

\title{Reciprocal symmetry and KNO scaling violation in proton-proton collisions}

\author{Mustapha Ouchen}
\email{mustapha.ouchen@ariel.ac.il}
\affiliation{Department of Physics, Ariel University, Ariel 4077601, Israel}

\author{Alex Prygarin}
\email{alexanderpr@ariel.ac.il}
\affiliation{Department of Physics, Ariel University, Ariel 4077601, Israel}

\begin{abstract}
We analyze the charged particle multiplicity distributions in $p-p$ collisions
and discuss the violation of the Koba--Nielsen--Olesen (KNO) scaling. We
extract the deviations from the leading exponential behavior of the KNO scaled
probability and identify a reciprocal symmetry $z\leftrightarrow 1/z$ in the
KNO violating corrections observed in the ATLAS and CMS data at
$\sqrt{s}=7,\,8,\,13$~TeV. The symmetry imposes a local constraint on the
multiplicity distribution at $n=\langle n\rangle$, namely
$P'(\langle n\rangle)=-P(\langle n\rangle)/\langle n\rangle$, which we verify
directly in the data. We use this constraint to extract the entanglement
entropy from the well-measured region $n\simeq\langle n\rangle$, avoiding the
large uncertainties associated with the distribution tail.
\end{abstract}
\maketitle

 \section{Introduction}
 
 Charged particle production in high-energy collisions provides insight into
the underlying QCD dynamics. The KNO scaling hypothesis~\cite{KNO,Polyakov}
states that at high energies the multiplicity distribution depends only on
the scaled variable $z=n/\langle n\rangle$. The experimental data for $p-p$
collisions from ATLAS~\cite{ATLAS} (re-analyzed in~\cite{Kulch}) and
CMS~\cite{CMS} show clear violations of this scaling.

 Charged particle multiplicity distributions have recently been studied in
the framework of quantum entanglement, following the proposal of Kharzeev
and   Levin~\cite{KharzeevLevin,Kharzeev2021} (see also~\cite{Kharzeev}) that
the partonic state probed in deep inelastic scattering (DIS) is a maximally
entangled state with entropy related to final-state hadron multiplicity.
This proposal has been examined through analyses of entanglement at
subnucleonic scales~\cite{TuKharzeevUllrich}, the H1 measurement~\cite{H1},
BFKL-based comparisons~\cite{HentschinskiKutak2022,HKStraka}, diffractive
DIS~\cite{HKKT2023}, QCD evolution~\cite{HKKT2024}, Monte Carlo
studies~\cite{HentschinskiJungKutak}, and dipole cascade
models~\cite{KutakPraszalowicz,KutakLokos}. Several recent works have
proposed alternatives or   extensions, including a diffusion-scaling
framework for high-multiplicity events~\cite{MoriggiNavarraMachado},
dipole evolution at high energy~\cite{LiuNowakZahed}, and a Gaussian
color-charge action at saturation~\cite{Dumitru}.

In the present paper we identify a reciprocal symmetry $f_s(z)=f_s(1/z)$ in
the KNO violating term, where $f_s(z)$ is defined as the relative deviation
of $\langle n\rangle P_n$ from the leading exponential $e^{-z}$. The symmetry
holds in the range $1/3<z<3$ for the ATLAS and CMS data at
$\sqrt{s}=7,\,8,\,13$~TeV. As a direct consequence of the symmetry, the
derivative of $f_s(z)$ vanishes at $z=1$, which translates into a local
relation $P'(\langle n\rangle)=-P(\langle n\rangle)/\langle n\rangle$ between
the value of the multiplicity distribution and its derivative at
$n=\langle n\rangle$. We verify this relation in the data at the few-percent
level and use it to extract the entanglement entropy from the well-measured
region $n\simeq\langle n\rangle$, avoiding the large uncertainties associated
with the distribution tail.

\section{Deviation from the exponential distribution}
 
The probability $P_n$ describes the multiplicity distribution of the produced
charged particles. The KNO scaling hypothesis~\cite{KNO,Polyakov} states that
at high energies the probability is given by
\begin{eqnarray}
P_n=\frac{1}{\langle n\rangle}\,\psi(z),
\end{eqnarray}
where $\psi(z)$ is a function of only one variable $z=n/\langle n\rangle$.
Deviations from this simple scaling form are usually referred to as KNO
scaling violation. The theoretical models and the experimental data for
$p-p$ collisions by ATLAS and CMS show the leading exponential behavior of
the scaled probability
\begin{eqnarray}
\langle n\rangle P_n\simeq e^{-z}\left(1+\mathcal{O}\!\left(\frac{1}{\langle n\rangle}\right)\right),
\end{eqnarray}
where $z=n/\langle n\rangle$ is the scaled variable that denotes the
deviation from the average $\langle n\rangle$.

\begin{figure}[!htb]
    \centering
    \begin{subfigure}[b]{0.49\columnwidth}
        \centering
        \includegraphics[width=\textwidth]{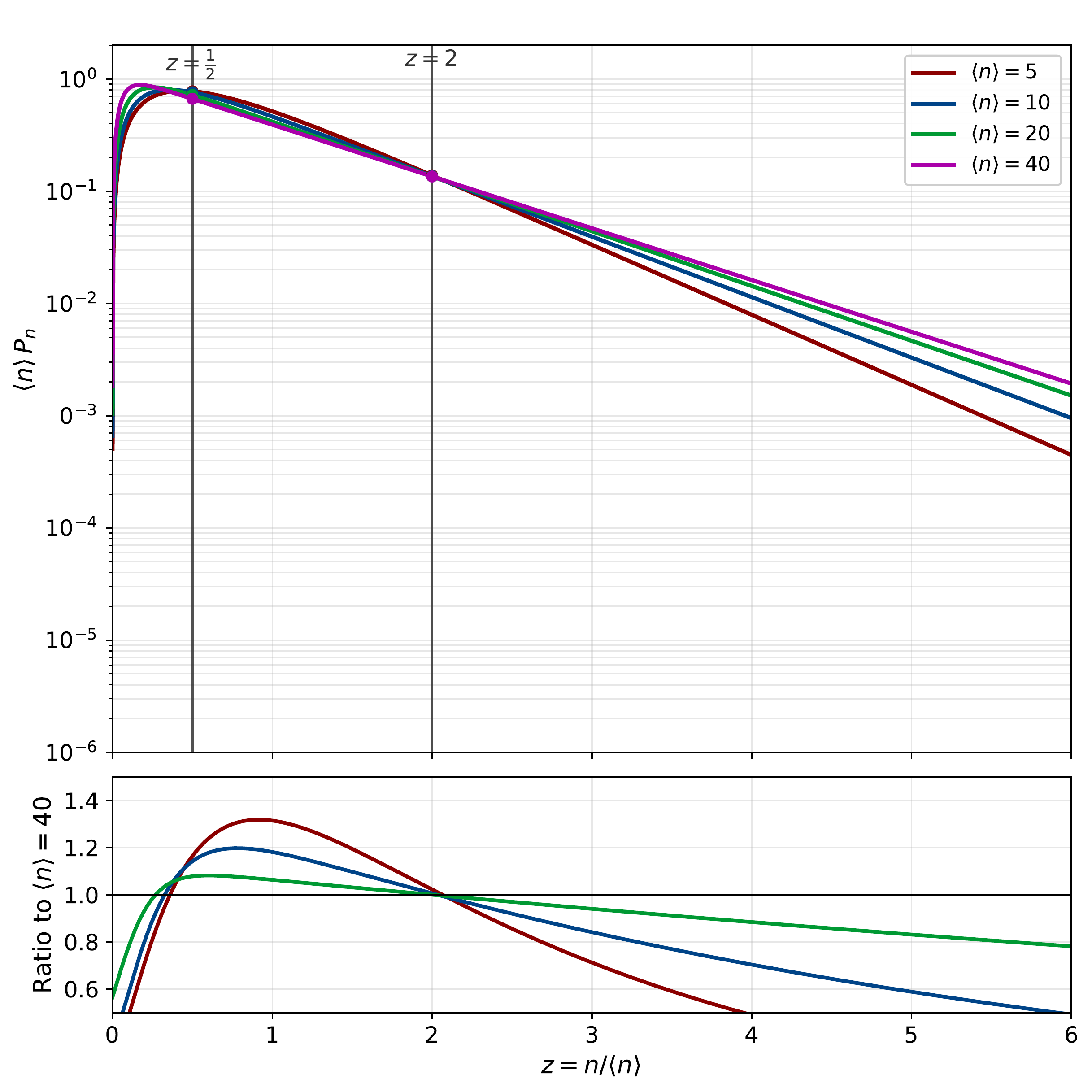}
        \caption{AGK model}
        \label{fig:agkcrossing}
    \end{subfigure}\hfill
    \begin{subfigure}[b]{0.49\columnwidth}
        \centering
        \includegraphics[width=\textwidth]{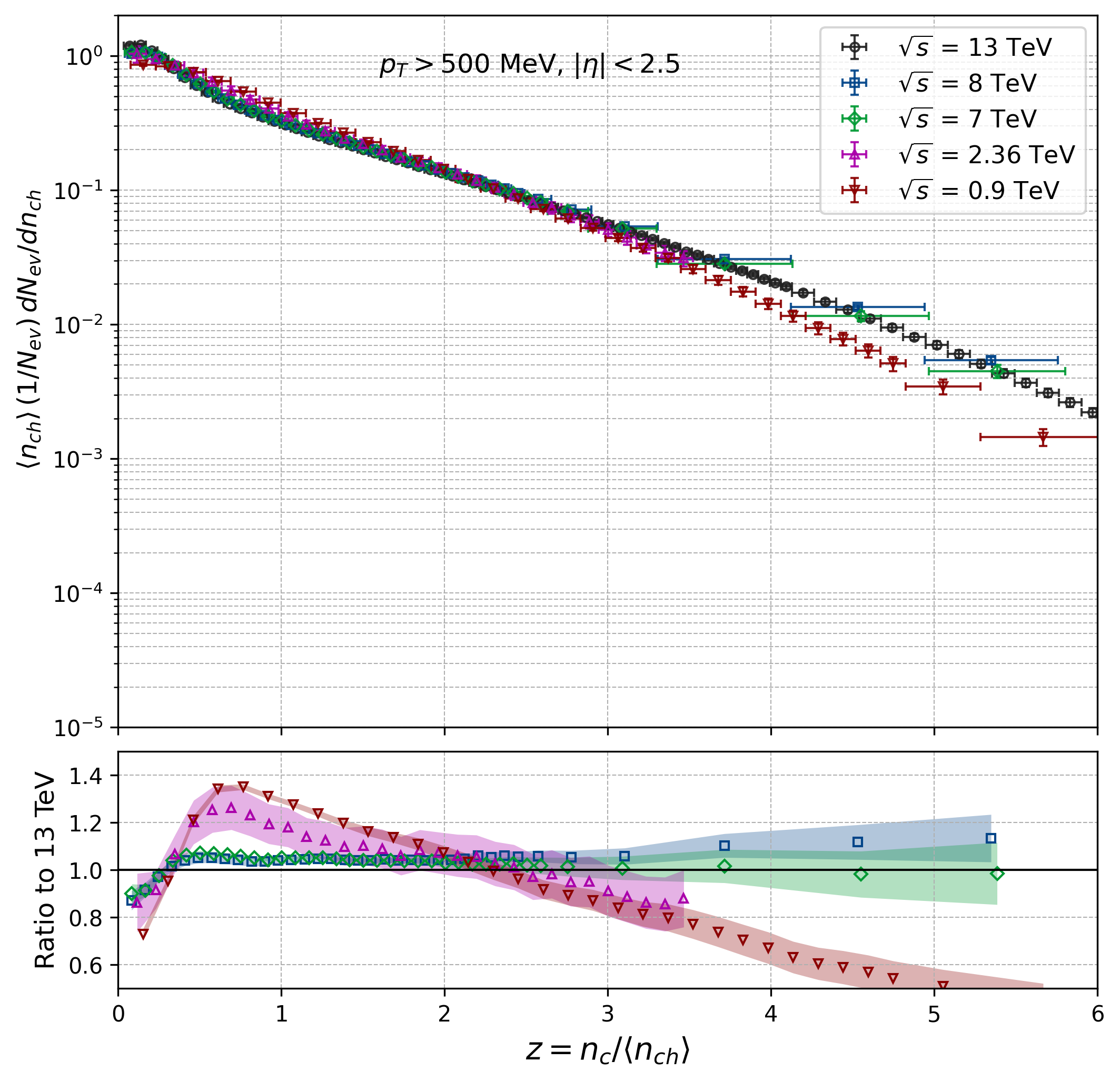}
        \caption{ATLAS data}
        \label{fig:atlascrossing}
    \end{subfigure}
    \caption{The figure shows the plot of $\langle n\rangle P_n$ as a function of the scaled variable $z=n/\langle n\rangle$ for the AGK model~(a) and the corresponding plot of the ATLAS experimental data~(b) for the charged multiplicity distribution at $p-p$ collisions at $\sqrt{s}=0.9,\,2.36,\,7,\,8,\,13$~TeV for $p_t>500$~MeV and $|\eta|<2.5$. In both plots we use the averages $\langle n\rangle$ calculated in~\cite{Kulch}. The intersection points $z=2$ and $z=1/2$ found in the AGK model are also clearly visible in the ATLAS experimental data.}
    \label{fig:crossing}
\end{figure}

It is useful to introduce a function $f_s(z)$, which measures the deviation
from the exponential distribution
\begin{eqnarray}
\langle n\rangle P_n=e^{-z}\left(1+f_s(z)\right).
\label{nPnfs}
\end{eqnarray}
If the function $f_s(z)$ is a function of only the variable $z$, then the
KNO scaling holds exactly. This is not the case for $p-p$ collisions, where
the experimental data show that the KNO scaling is violated because
$\langle n\rangle P_n$ versus $z$ curves are quite different for different
energies and other parameters as depicted in Fig.~\ref{fig:crossing}(b).

In the right plot of Fig.~\ref{fig:crossing} we used the averages
$\langle n\rangle$ calculated in~\cite{Kulch} that are listed in
Table~\ref{table:average}.

\begin{table}[htbp]
    \centering
    \renewcommand{\arraystretch}{1.3}
    \begin{tabular}{cc}
        \hline\hline
        $\sqrt{s}$, [TeV] & $\langle n\rangle$ \\[1ex]
        \hline
        13   & $14.66\pm 0.04$ \\
        8    & $12.25\pm 0.03$ \\
        7    & $11.98\pm 0.05$ \\
        2.36 & $8.66\pm 0.51$ \\
        0.9  & $6.53\pm 0.03$ \\
        \hline\hline
    \end{tabular}
    \caption{Averages of the charged multiplicity distribution in $p-p$ collisions at different energies for $p_T^{\min}>500$~MeV and $|\eta|<2.5$ as calculated in~\cite{Kulch}.}
    \label{table:average}
\end{table}

 The probability $P_n$ was calculated by Mueller~\cite{Mueller} for the color
dipole model  using a Markov chain equation, and was later used by Kharzeev
and Levin~\cite{KharzeevLevin} for studying entanglement in $p-p$ scattering.
The Mueller--Kharzeev--Levin (MKL) model is a pure geometric   distribution
given by 
\begin{eqnarray}
P^{\mathrm{Mueller}}_n=e^{-\alpha Y}\,(1-e^{-\alpha Y})^{n-1},
\end{eqnarray}
where $\alpha$ and $Y$ are the coupling constant and the rapidity
respectively.

An extended version of the MKL model was developed by the
authors~\cite{OuchenPrygarin1} by  implementing the  relative weights found
by Abramovsky, Gribov, and Kancheli (AGK)~\cite{AGK} in the Markov chain
evolution equation. To avoid the negative probabilities arising from the
negative AGK weights in Minkowski space, the evolution was reformulated
in Euclidean space~\cite{OuchenPrygarin1}. The resulting AGK model is
slightly different from the pure geometric distribution
\begin{eqnarray}
P^{\mathrm{AGK}}_n&=&\left(1-\frac{2^n}{3^n}\right)e^{-\alpha Y}\,(1-e^{-\alpha Y})^{n-1} \nonumber \\
&&\times\,(1+2e^{-\alpha Y}).
\end{eqnarray}

 The two models are very close to each other and numerically describe the
experimental data in a very similar way, even though the AGK model better
describes the experimental values for the moments
$C_q$~\cite{OuchenPrygarin1}. However, there is a qualitative difference
between the two models when it comes to the corrections to the leading
$e^{-z}$ behavior. Both models exhibit a universal fixed point of the
scaled probability $\langle n\rangle P_n$ at $z=2$, where curves at
different $\langle n\rangle$ intersect. As recently shown by the
authors~\cite{OuchenPrygarin2}, this fixed point is a generic feature of
any unitary model that approximates
 a memoryless distribution at high
energy, with KNO-violating corrections of order $1/\langle n\rangle^2$ near
$z=2$, reflecting the maximal entanglement of the final states. Besides
this, the AGK-based model reveals another approximate intersection point
at $z=1/2$, which is also present in the experimental data as shown in
Fig.~\ref{fig:crossing}(a). The intersection point $z=1/2$ is absent in
the MKL model and other popular models. The charged multiplicity
distribution ATLAS data for $p-p$ collisions show evidence for the
approximate intersection 
point at $z=1/2$ of the $\langle n\rangle P_n$
curves at different energies as illustrated in Fig.~\ref{fig:crossing}.
The two intersection points $z=1/2$ and $z=2$ are themselves reciprocal
partners, suggestive of a residual symmetry that becomes more pronounced
in the data, as we discuss next.

Moreover, the ATLAS data reveal an intriguing reciprocal symmetry
$z\leftrightarrow 1/z$ of the KNO violating term $f_s(z)$ in the range
$1/3<z<3$ for energies $\sqrt{s}=2.36,\,7,\,8,\,13$~TeV, as we show in the
next section.

 \section{Reciprocal symmetry $z\leftrightarrow 1/z$}
\label{sec:reciprocal}

It is convenient to rewrite the function $f_s(z)$ defined in (\ref{nPnfs})
as the relative deviation of the measured distribution from the leading
exponential distribution
\begin{eqnarray}
f_s(z)=\frac{\langle n\rangle P_n-e^{-z}}{e^{-z}}.
\label{fs}
\end{eqnarray}

 The function $f_s(z)$ generally depends on both $z$ and the average
$\langle n\rangle$; if it depended only on $z$, the KNO scaling would hold
exactly. The function $f_s(z)$ is constructed to measure the relative
deviation from the memoryless exponential behavior $e^{-z}$, i.e.\ from the
geometric (memoryless) distribution that arises in the color-dipole
picture~\cite{Baker,KharzeevLevin,TuKharzeevUllrich}.

\begin{figure}[!htb]
    \centering
    \begin{subfigure}[b]{0.49\columnwidth}
        \centering
        \includegraphics[width=\textwidth]{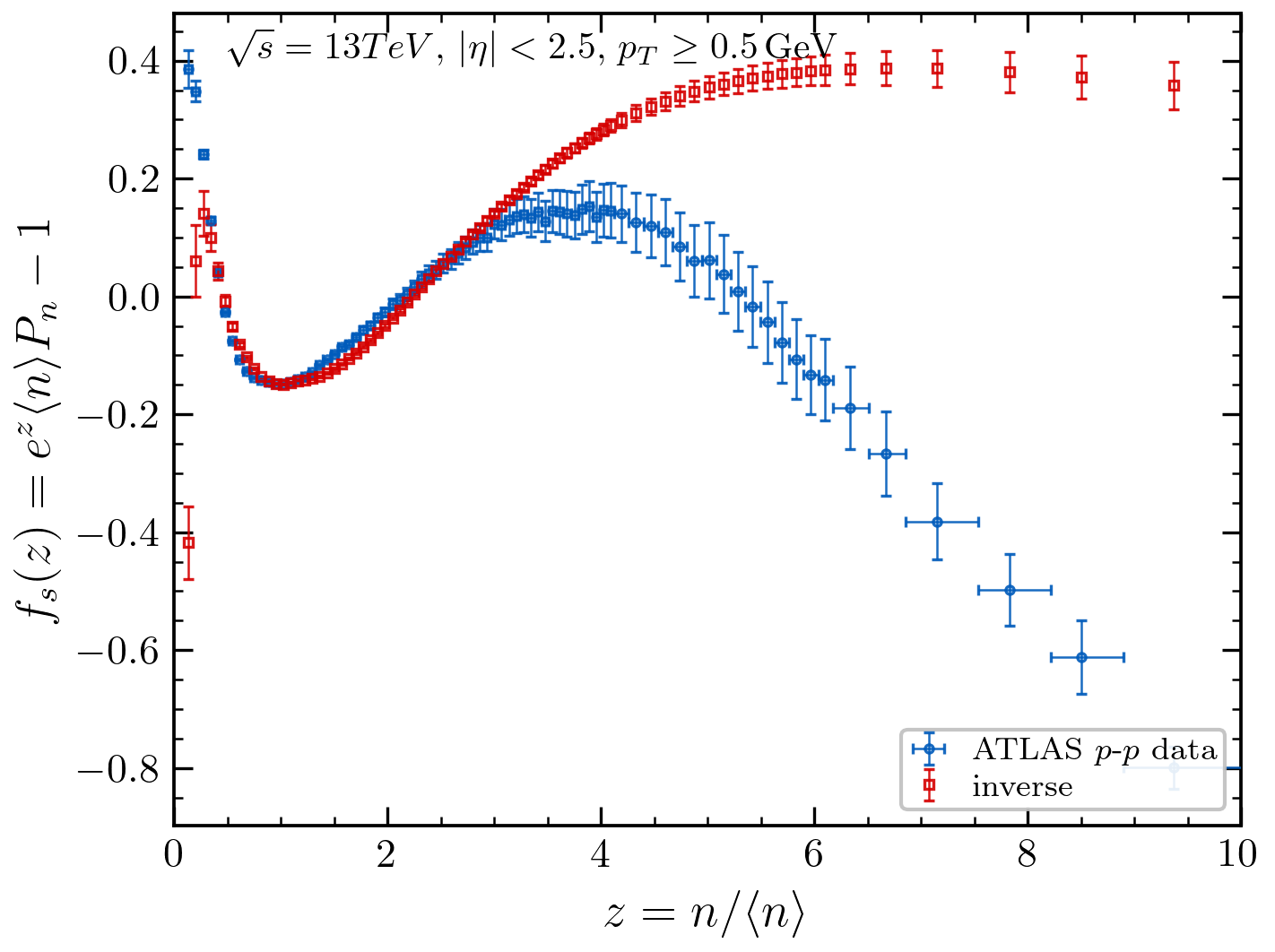}
        \caption{$\sqrt{s}=13$~TeV}
        \label{fig:fs13}
    \end{subfigure}\hfill
    \begin{subfigure}[b]{0.49\columnwidth}
        \centering
        \includegraphics[width=\textwidth]{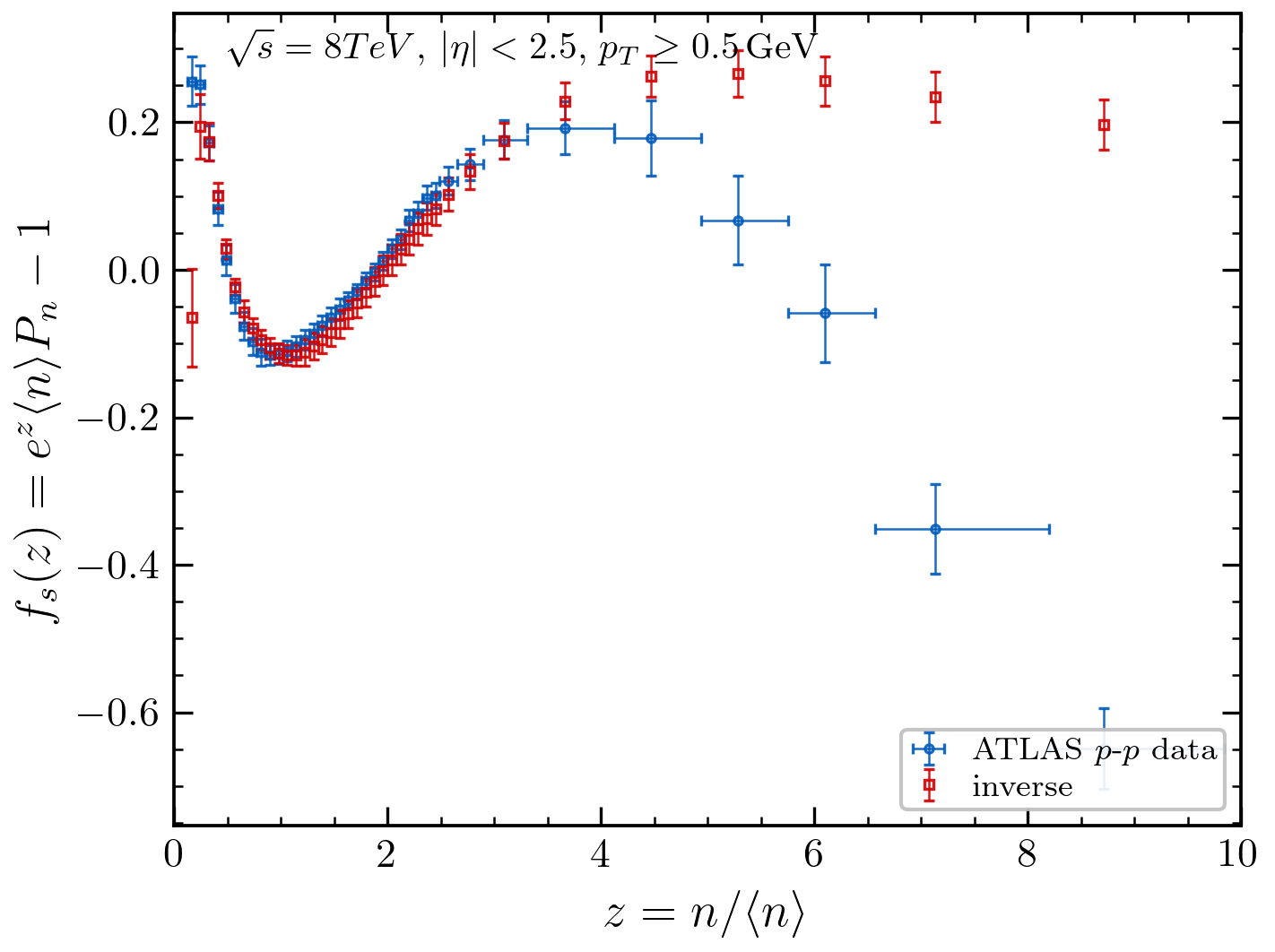}
        \caption{$\sqrt{s}=8$~TeV}
        \label{fig:fs8}
    \end{subfigure}

    \vspace{0.4cm}

    \begin{subfigure}[b]{0.49\columnwidth}
        \centering
        \includegraphics[width=\textwidth]{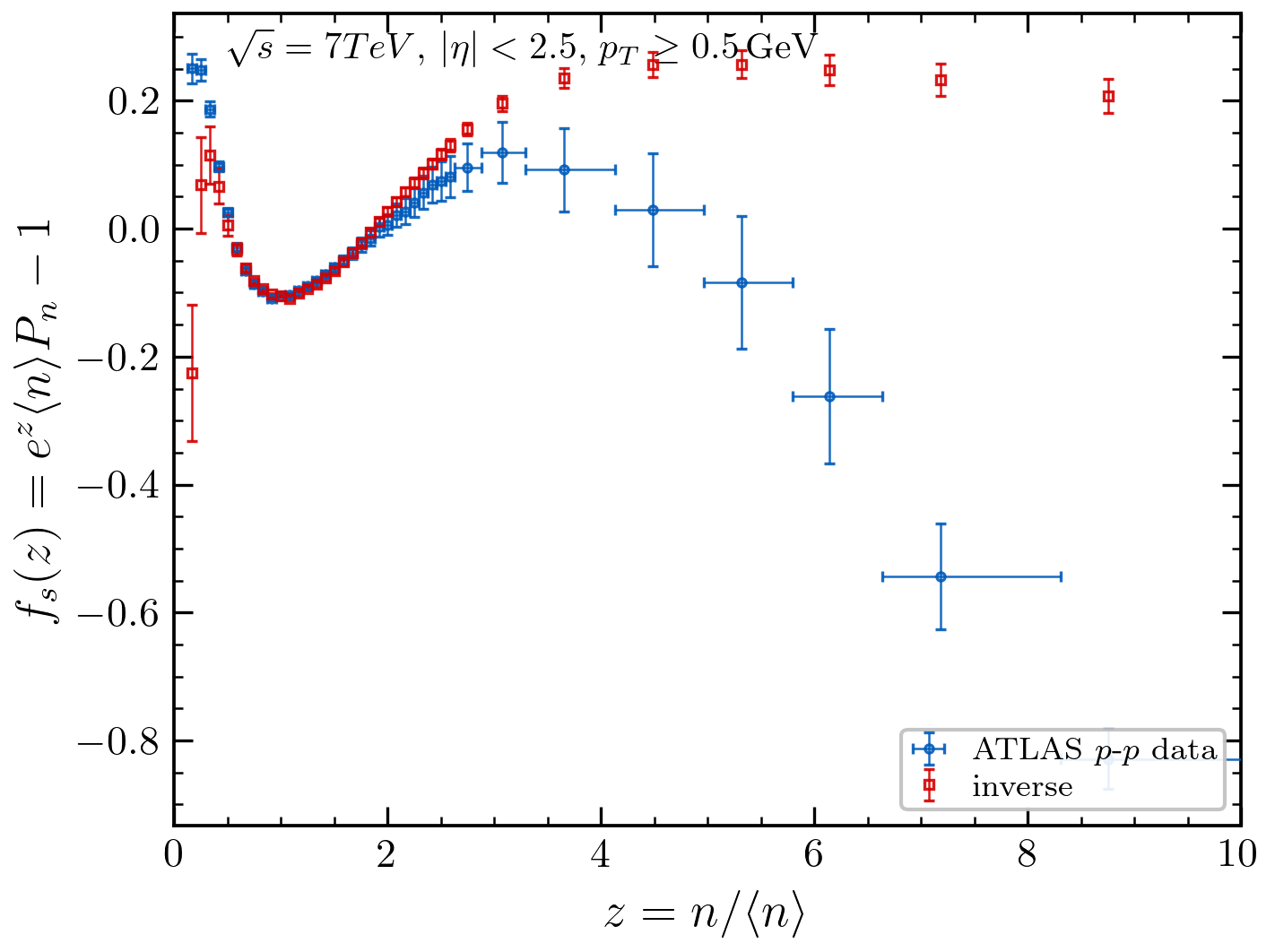}
        \caption{$\sqrt{s}=7$~TeV}
        \label{fig:fs7}
    \end{subfigure}\hfill
    \begin{subfigure}[b]{0.49\columnwidth}
        \centering
        \includegraphics[width=\textwidth]{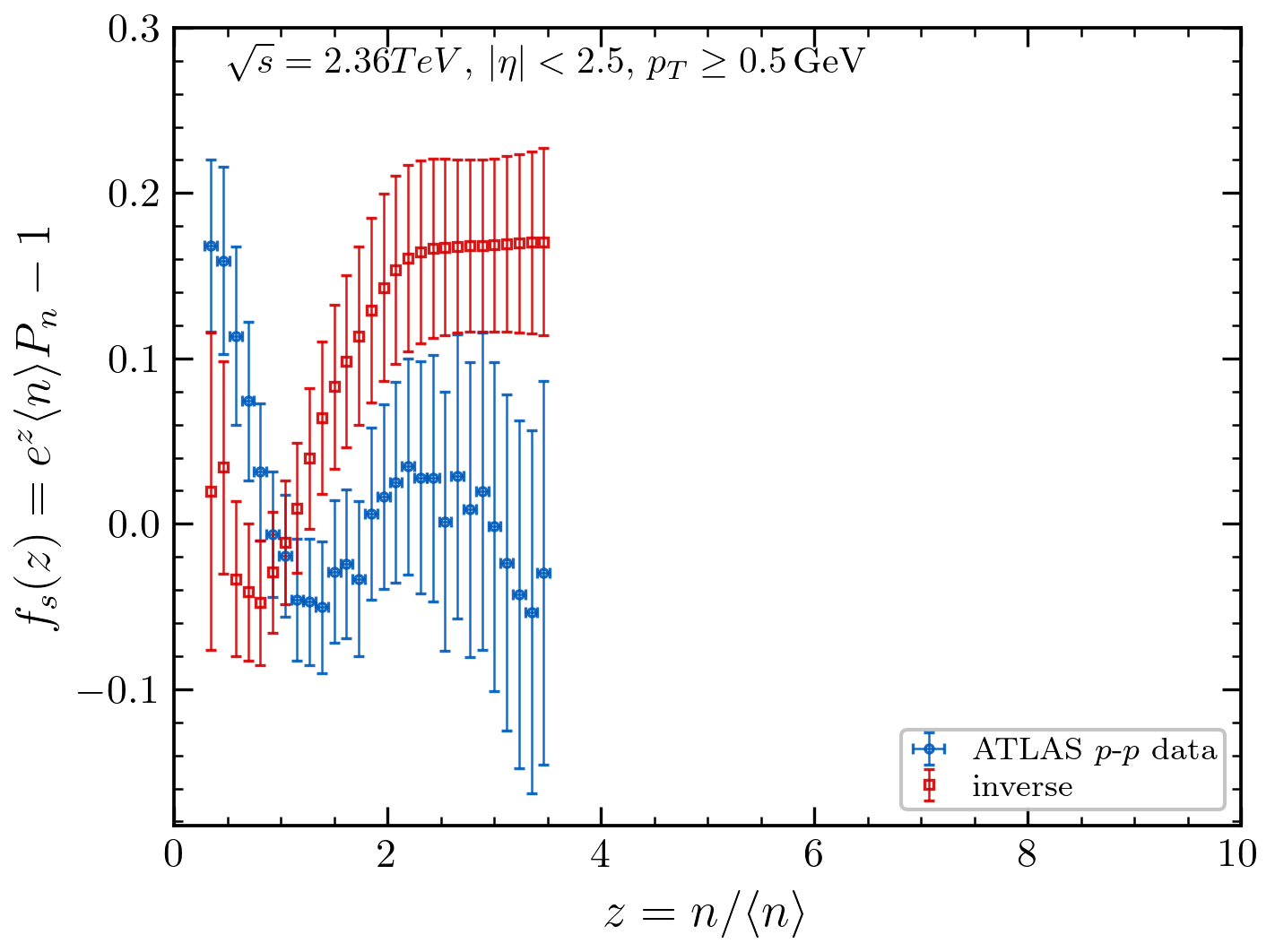}
        \caption{$\sqrt{s}=2.36$~TeV}
        \label{fig:fs236}
    \end{subfigure}

    \caption{The ATLAS data plots for $f_s(z)$ defined in (\ref{fs}) at $\sqrt{s}=2.36,\,7,\,8,\,13$~TeV, $p_t>500$~MeV and $|\eta|<2.5$. The blue points represent $f_s(z)$ as a function of $z$, and the red points represent $f_s(1/z)$ of the inverse variable $1/z$, illustrating the reciprocal symmetry $z\leftrightarrow 1/z$ in the range $1/3<z<3$ at $\sqrt{s}=7,\,8,\,13$~TeV. The averages $\langle n\rangle$ used are those calculated in~\cite{Kulch} and given in Table~\ref{table:average}. The plot at $\sqrt{s}=2.36$~TeV demonstrates that the reciprocal symmetry does not hold at lower energies.}
    \label{fig:fsreciprocal}
\end{figure}

   The ATLAS and CMS experimental data for the charged multiplicity
distribution at $\sqrt{s}=7$, $8$, and $13$~TeV show evidence for a
remarkable reciprocal symmetry of the KNO violating term $f_s(z)$. The
reciprocal symmetry $f_s(z)=f_s(1/z)$ holds well in a wide range of
fluctuations from the average $\langle n\rangle$, namely for $1/3<z<3$, as
shown in Fig.~\ref{fig:fsreciprocal} for the ATLAS data at
$\sqrt{s}=7$, $8$, and $13$~TeV. For $\sqrt{s}=2.36$~TeV the reciprocal
symmetry is much less pronounced. We include the $\sqrt{s}=2.36$~TeV plot
to illustrate that the symmetry builds up with energy.

\subsection{Gaussian parametrization}

The reciprocal symmetry $f_s(z)=f_s(1/z)$ is automatically respected by any
even function of $\ln z$, since $\ln(1/z)=-\ln z$. The simplest such
function with finite support around $z=1$ is a Gaussian in $\ln z$, which
we adopt as a convenient parametrization
\begin{eqnarray}
f^{\mathrm{fit}}_s(z)=a+b\,e^{-c(\ln z-\mu)^2},
\label{fsgauss}
\end{eqnarray}
    where $\mu$ measures the departure from exact reciprocal symmetry: $\mu=0$
corresponds to the exactly symmetric case. The Gaussian in $\ln z$ serves
here as a convenient phenomenological parametrization that makes the
symmetry manifest; any even function of $\ln z$ would respect the symmetry
equally. The dynamical origin of both the symmetry itself and the specific
shape of $f_s(z)$ is left for future investigation.

\begin{figure}[!htb]
    \centering
    \begin{subfigure}[b]{0.49\columnwidth}
        \centering
        \includegraphics[width=\textwidth]{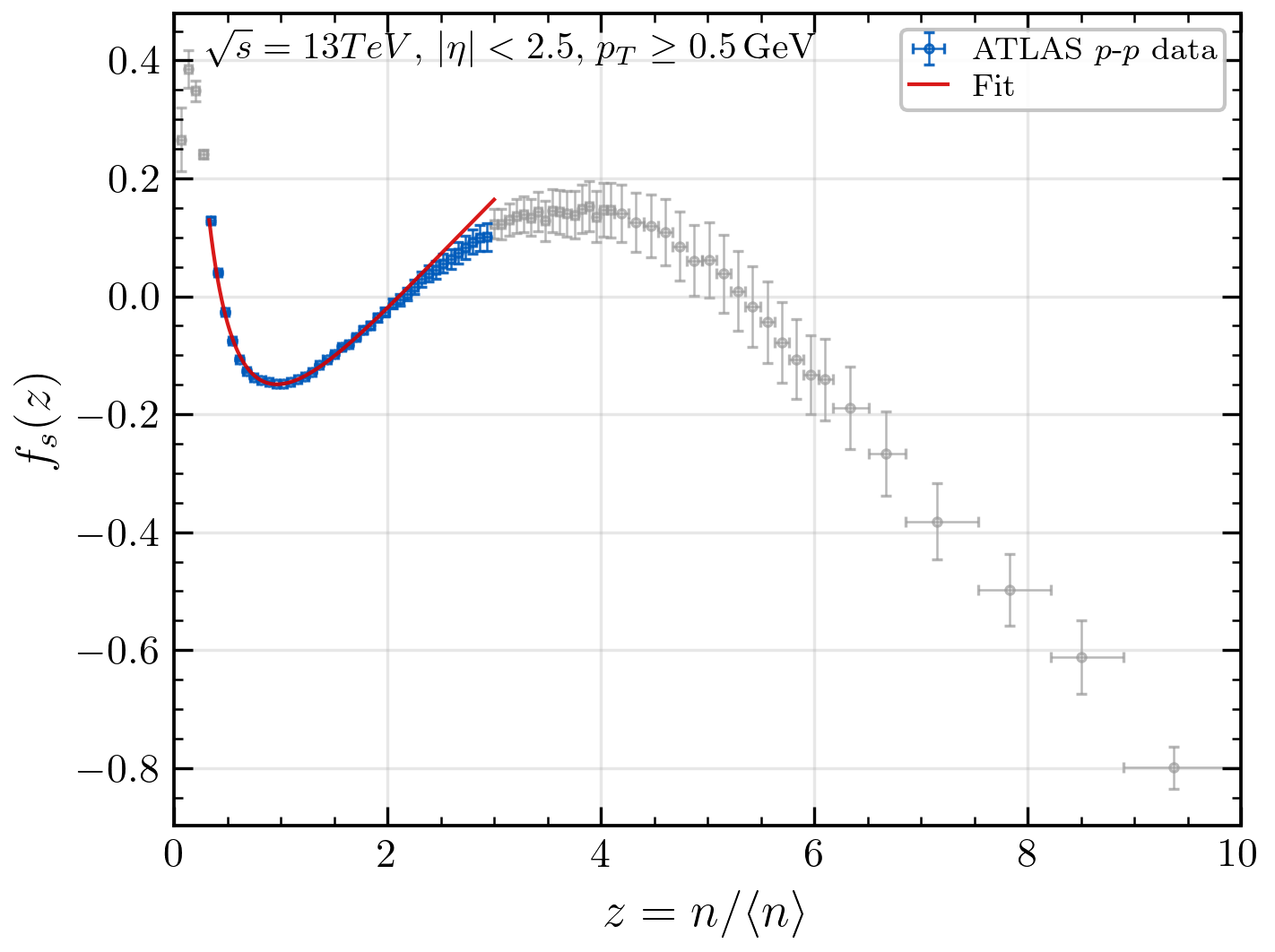}
        \caption{$\sqrt{s}=13$~TeV}
        \label{fig:fsg13}
    \end{subfigure}\hfill
    \begin{subfigure}[b]{0.49\columnwidth}
        \centering
        \includegraphics[width=\textwidth]{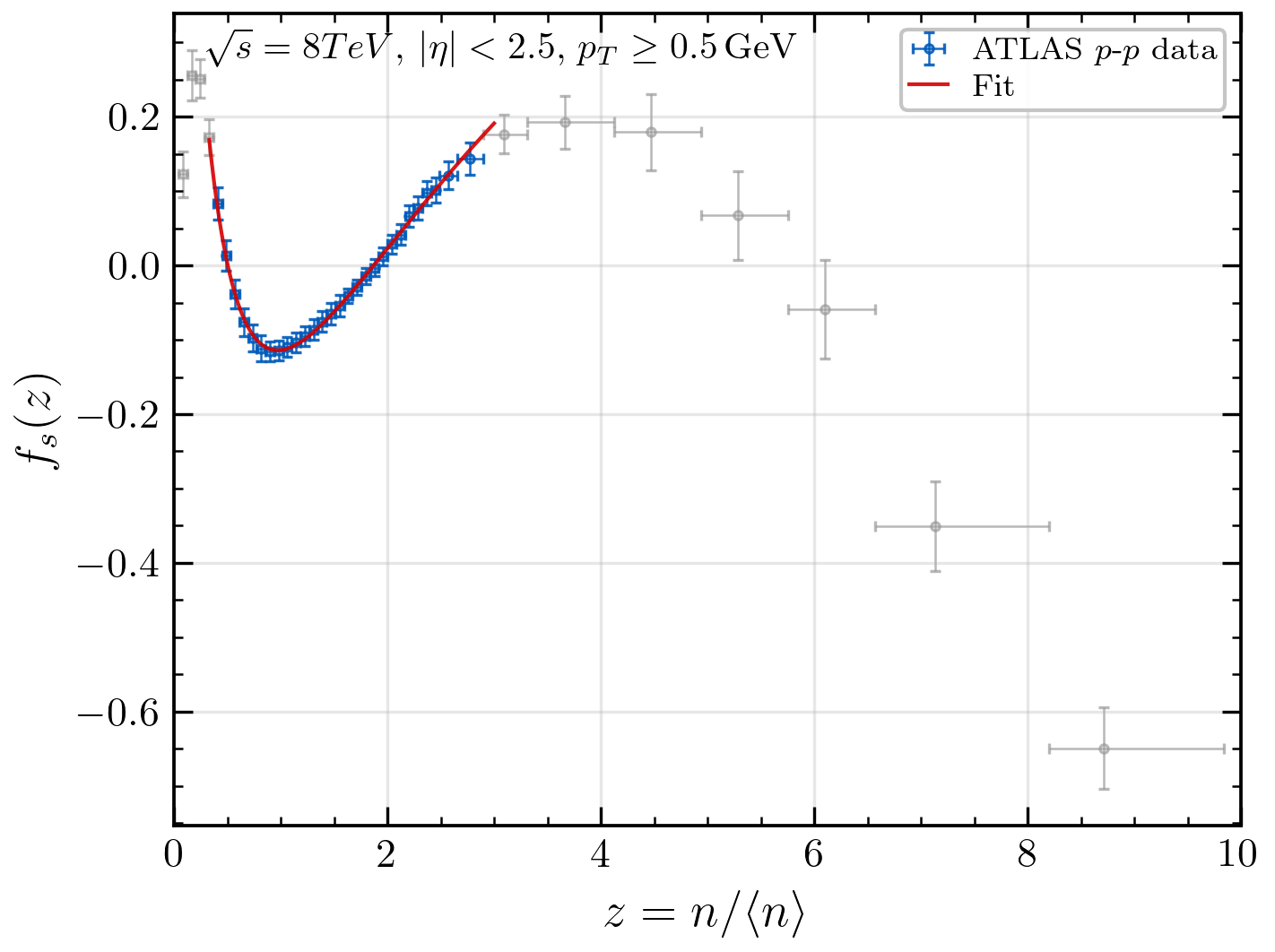}
        \caption{$\sqrt{s}=8$~TeV}
        \label{fig:fsg8}
    \end{subfigure}

    \vspace{0.4cm}

    \begin{subfigure}[b]{0.49\columnwidth}
        \centering
        \includegraphics[width=\textwidth]{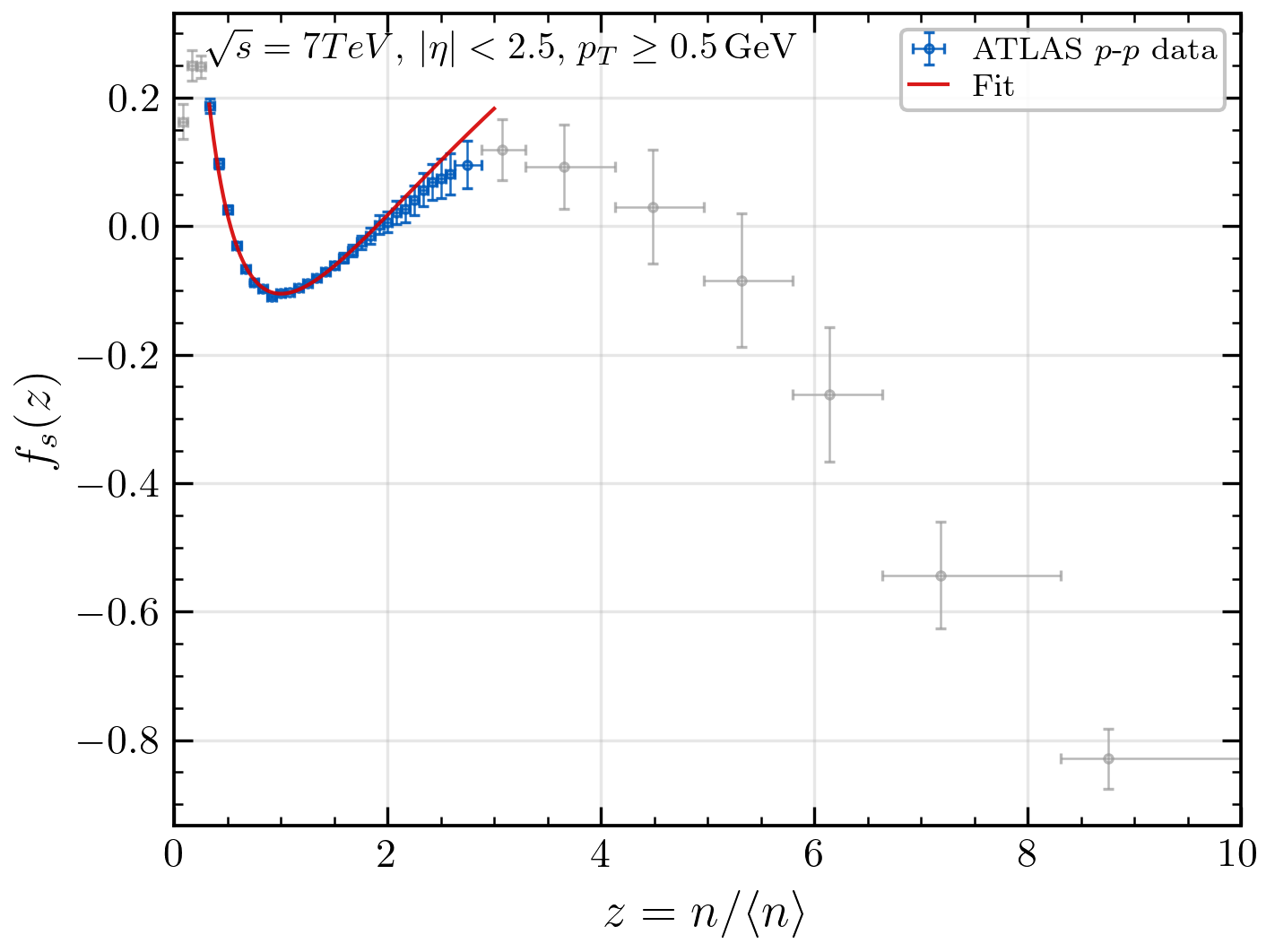}
        \caption{$\sqrt{s}=7$~TeV}
        \label{fig:fsg7}
    \end{subfigure}\hfill
    \begin{subfigure}[b]{0.49\columnwidth}
        \centering
        \includegraphics[width=\textwidth]{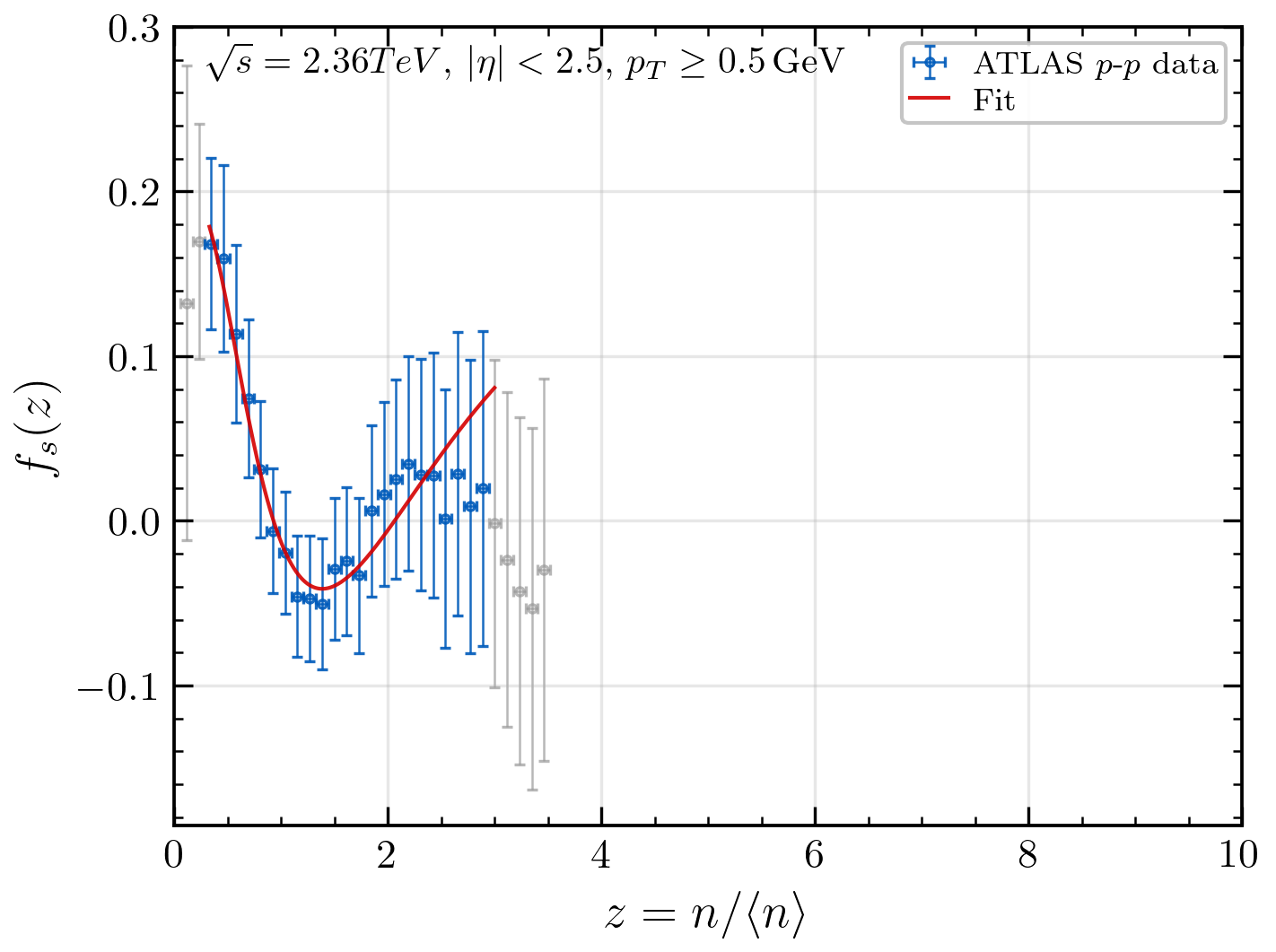}
        \caption{$\sqrt{s}=2.36$~TeV}
        \label{fig:fsg236}
    \end{subfigure}

    \caption{Gaussian fit (red line) of the deviation function $f_s(z)$ defined in (\ref{fs}) against $z$ for the ATLAS data (blue points) for $p-p$ collisions at $\sqrt{s}=2.36,\,7,\,8,\,13$~TeV, $p_t>500$~MeV and $|\eta|<2.5$. The Gaussian of $\ln z$ given by (\ref{fsgauss}) shows good agreement with the experimental data and reflects the reciprocal symmetry $z\leftrightarrow 1/z$ in the range $1/3<z<3$ for $\sqrt{s}=7,\,8,\,13$~TeV. The plot at $\sqrt{s}=2.36$~TeV is presented to illustrate that the reciprocal symmetry is less pronounced at lower energies and builds up at higher energies. The grey dots represent data points beyond the fitting range.}
    \label{fig:fsgauss}
\end{figure}

\begin{figure}[!htb]
    \centering
    \begin{subfigure}[b]{0.49\columnwidth}
        \centering
        \includegraphics[width=\textwidth]{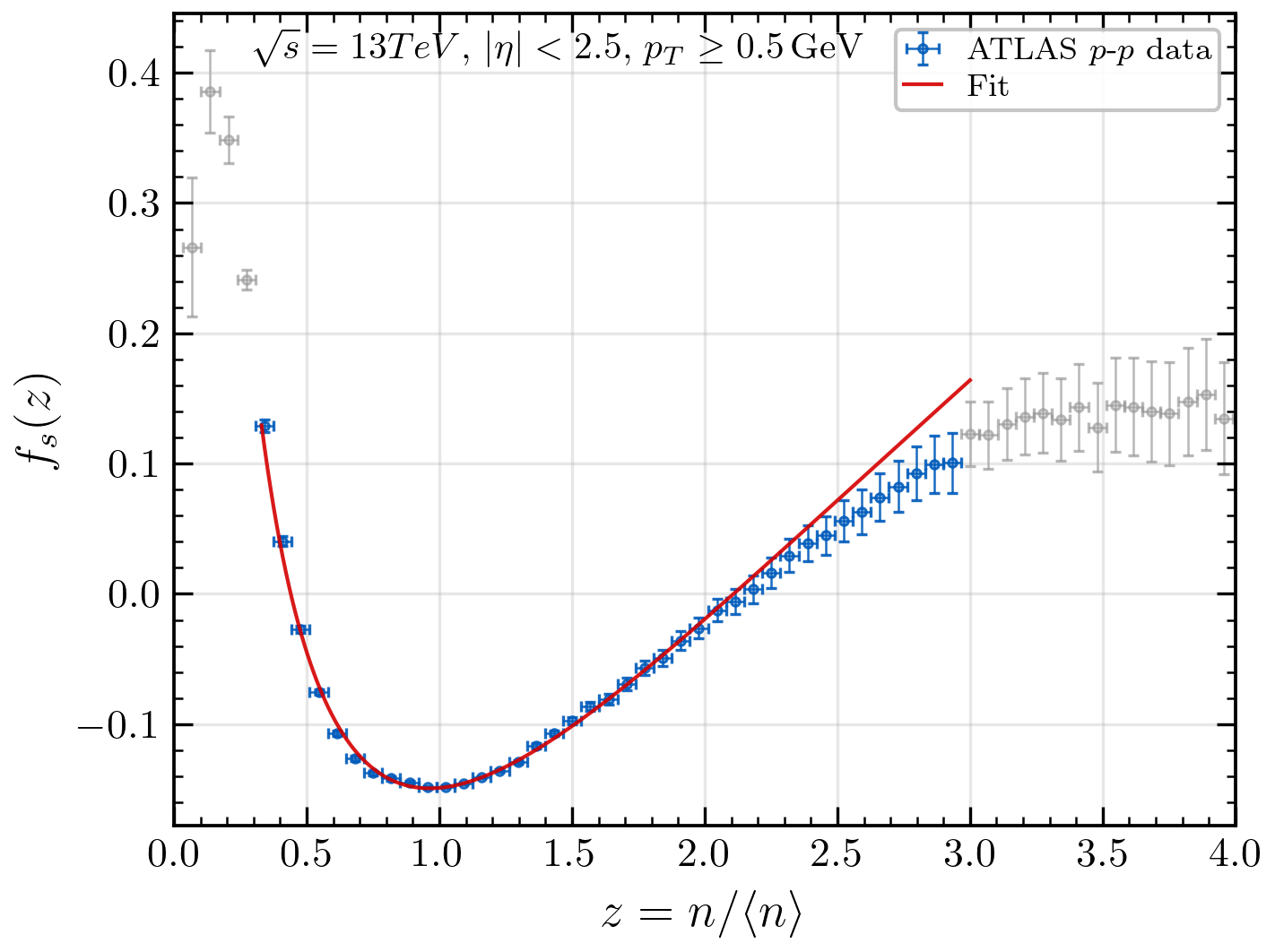}
        \caption{$\sqrt{s}=13$~TeV}
        \label{fig:fszg13}
    \end{subfigure}\hfill
    \begin{subfigure}[b]{0.49\columnwidth}
        \centering
        \includegraphics[width=\textwidth]{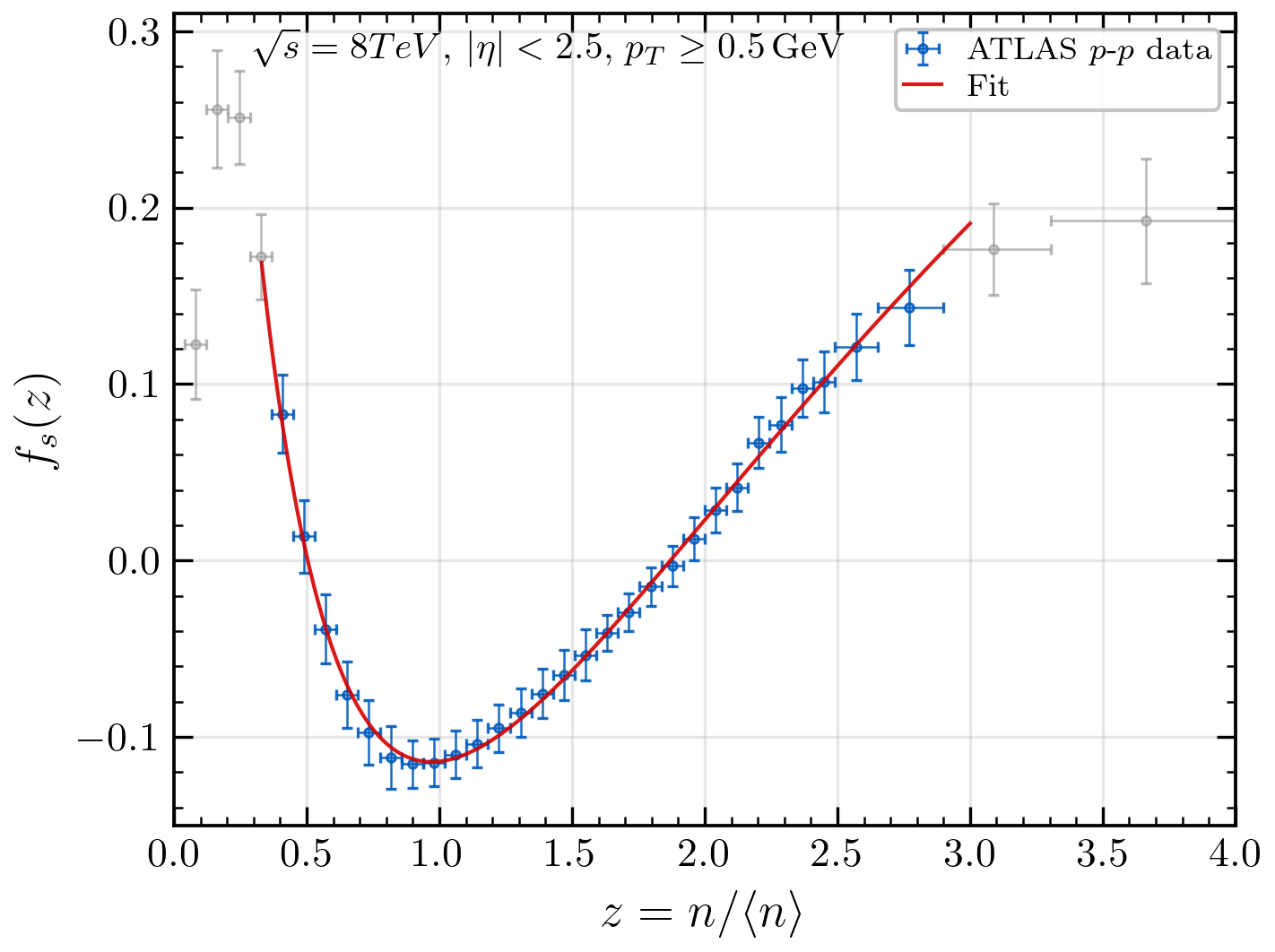}
        \caption{$\sqrt{s}=8$~TeV}
        \label{fig:fszg8}
    \end{subfigure}

    \vspace{0.4cm}

    \begin{subfigure}[b]{0.49\columnwidth}
        \centering
        \includegraphics[width=\textwidth]{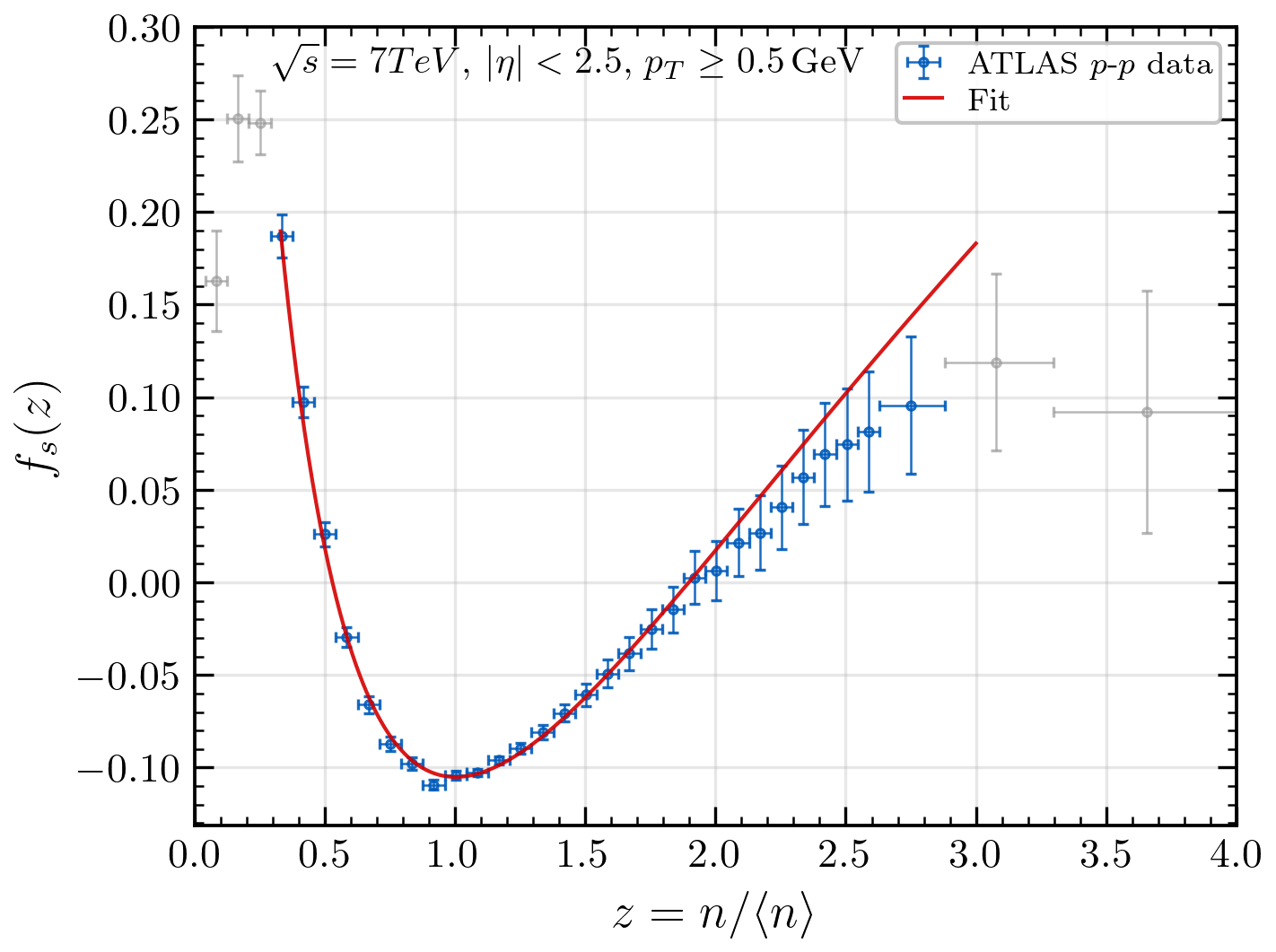}
        \caption{$\sqrt{s}=7$~TeV}
        \label{fig:fszg7}
    \end{subfigure}\hfill
    \begin{subfigure}[b]{0.49\columnwidth}
        \centering
        \includegraphics[width=\textwidth]{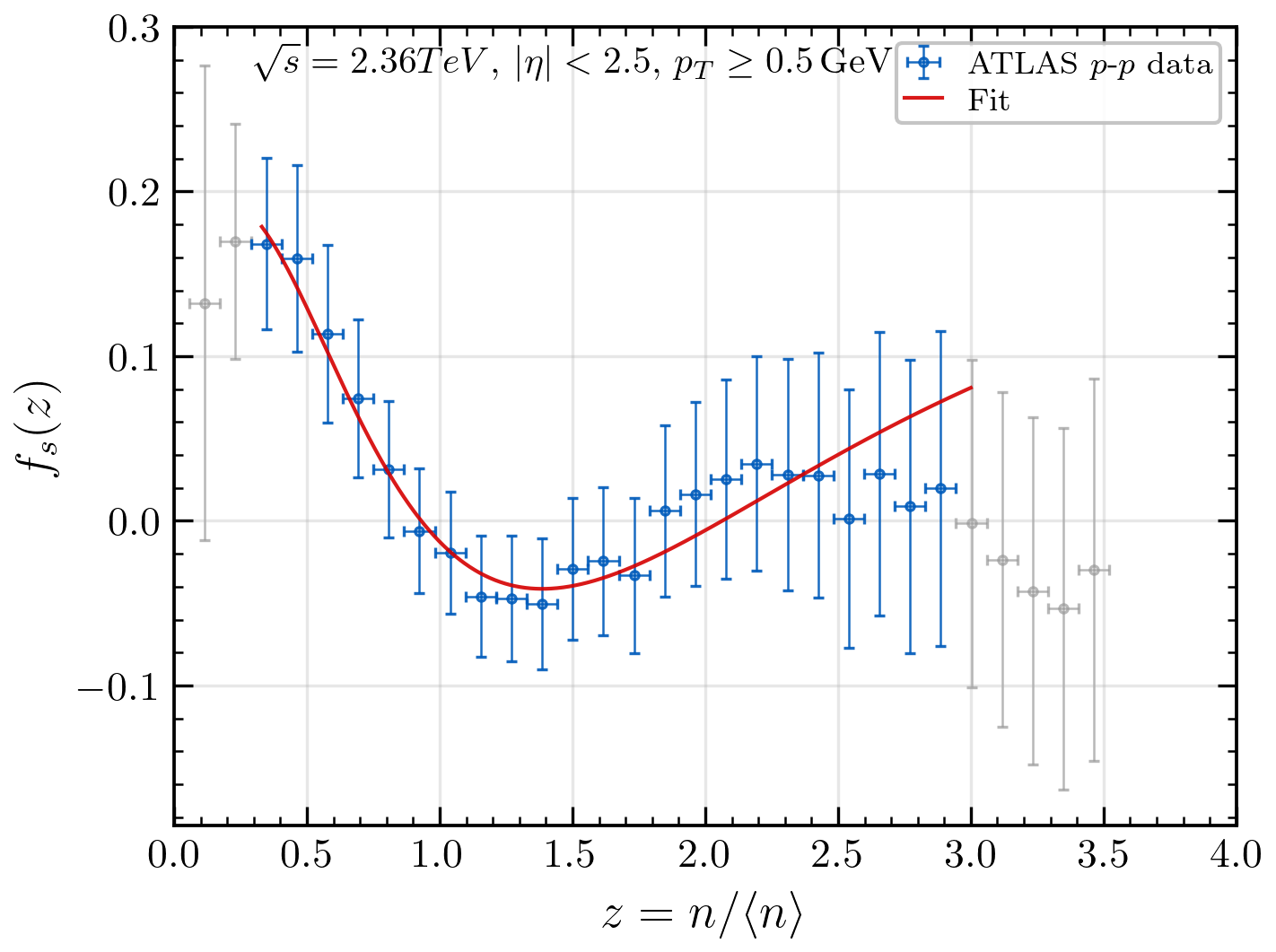}
        \caption{$\sqrt{s}=2.36$~TeV}
        \label{fig:fszg236}
    \end{subfigure}

    \caption{Zoomed Gaussian fit (red line) for $z\le 4$ of the deviation function $f_s(z)$ defined in (\ref{fs}) for the ATLAS data (blue points) for $p-p$ collisions at $\sqrt{s}=2.36,\,7,\,8,\,13$~TeV, $p_t>500$~MeV and $|\eta|<2.5$. The Gaussian of $\ln z$ given by (\ref{fsgauss}) shows good agreement with the experimental data and reflects the reciprocal symmetry $z\leftrightarrow 1/z$ in the range $1/3<z<3$ for $\sqrt{s}=7,\,8,\,13$~TeV. The plot at $\sqrt{s}=2.36$~TeV illustrates that the reciprocal symmetry is less pronounced at lower energies and builds up at higher energies. The grey dots represent data points beyond the fitting range.}
    \label{fig:zoomedfsgauss}
\end{figure}

    The fit results in the range $1/3<z<3$ are shown in
Table~\ref{tab:gaussparameters} and compared to the data in
Figs.~\ref{fig:fsgauss} and~\ref{fig:zoomedfsgauss}. The smallness of the
shift parameter $\mu$ for $\sqrt{s}=7,\,8,\,13$~TeV reflects the fact that
the reciprocal symmetry holds well at those energies, while at
$\sqrt{s}=2.36$~TeV $|\mu|$ is much larger because the symmetry is much
less pronounced.

 \begin{table*}[ht]
\centering
\renewcommand{\arraystretch}{1.3}
\begin{tabular}{ccccc}
\hline\hline
\textbf{Energy ($\sqrt{s}$)} & $\boldsymbol{a}$ & $\boldsymbol{b}$ & $\boldsymbol{c}$ & $\boldsymbol{\mu}$ \\
\hline
13 TeV & $0.59623\pm 0.15575$ & $-0.74619\pm 0.15432$ & $0.36615\pm 0.09314$ & $-0.01574\pm 0.00549$ \\
8 TeV  & $0.71888\pm 0.15820$ & $-0.83440\pm 0.15721$ & $0.35141\pm 0.07807$ & $-0.02889\pm 0.00401$ \\
7 TeV  & $0.76508\pm 0.31955$ & $-0.86774\pm 0.31800$ & $0.30193\pm 0.12938$ & $\phantom{-}0.03357\pm 0.00702$ \\
\hline\hline
\end{tabular}
\caption{Best-fit parameters for $f^{\mathrm{fit}}_s(z)=a+b\,e^{-c(\ln z-\mu)^2}$ restricted to the range $1/3<z<3$. The fitted function is compared to the corresponding experimental curves in Fig.~\ref{fig:fsgauss} and Fig.~\ref{fig:zoomedfsgauss}.}
\label{tab:gaussparameters}
\end{table*}

   The reciprocal symmetry is not observed outside the range $1/3<z<3$,
suggesting that the dynamics governing the tails of the distribution
differ from those near the average $\langle n\rangle$. Recent
work~\cite{MoriggiNavarraMachado} has proposed that high-multiplicity
events depart from  
 KNO scaling and are better described by a
diffusion-scaling framework, consistent with the picture that the tail and
the central region are governed by distinct dynamical regimes. This is
also supported by the fact that theory reproduces the experimental
moments $C_2$ and $C_3$ much better than $C_4$ and $C_5$, which
capture the tail of the multiplicity distribution.

Possible candidates for the additional dynamics in the tail
include multi-parton interactions and collective motion. Our AGK-based
model exhibits the intersection points at $z=1/2$ and $z=2$ but does not
exhibit the full reciprocal symmetry, which we attribute to the absence of
Pomeron loops in the model (i.e.\ the omission of dipole merging terms in
the color-dipole language).

\subsection{Reciprocal transformation of $\langle n\rangle P_n$}

\begin{figure}[!htb]
    \centering
    \begin{subfigure}[b]{0.49\columnwidth}
        \centering
        \includegraphics[width=\textwidth]{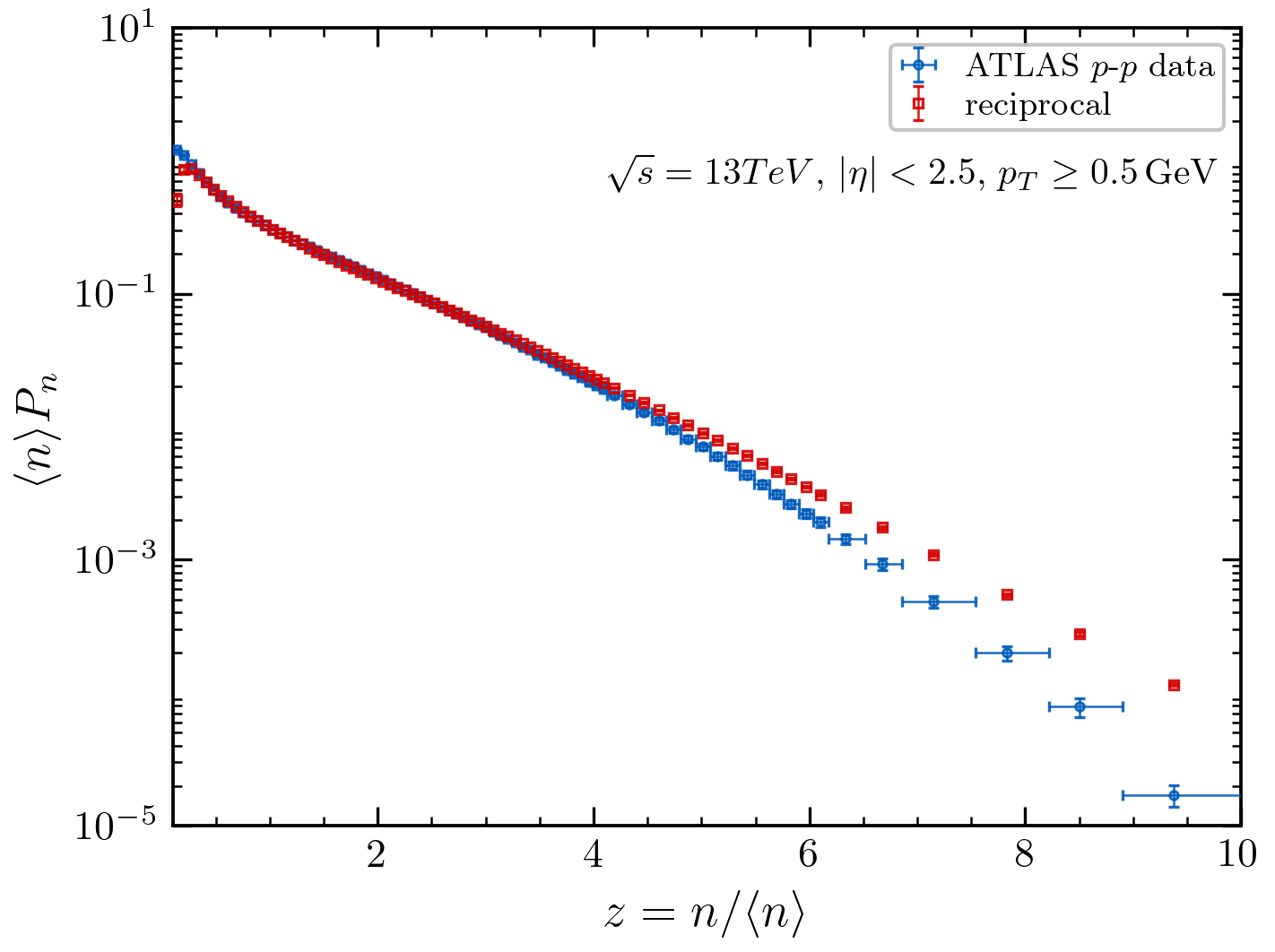}
        \caption{$\sqrt{s}=13$~TeV}
        \label{fig:psi13}
    \end{subfigure}\hfill
    \begin{subfigure}[b]{0.49\columnwidth}
        \centering
        \includegraphics[width=\textwidth]{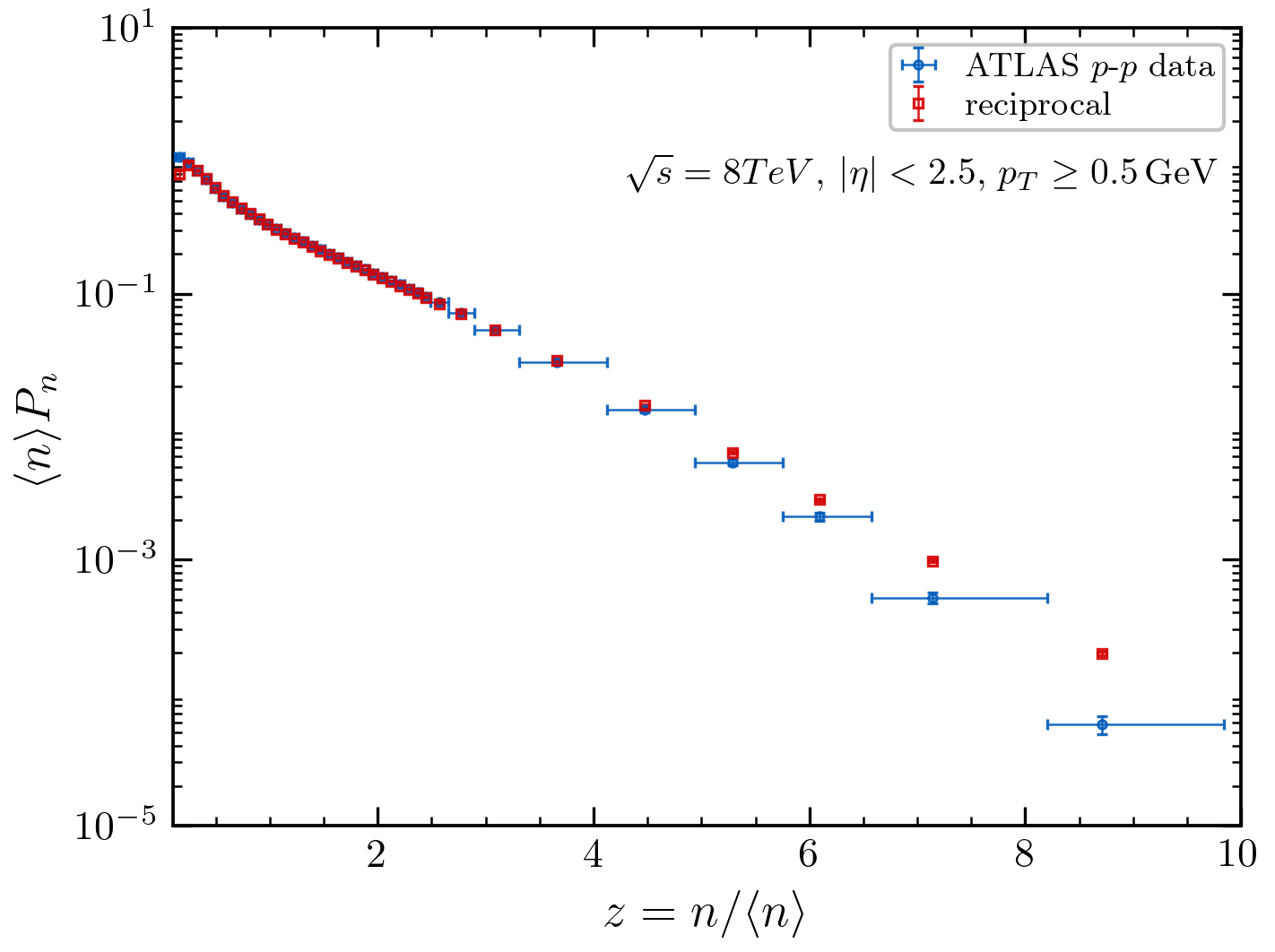}
        \caption{$\sqrt{s}=8$~TeV}
        \label{fig:psi8}
    \end{subfigure}
    \vspace{0.4cm}

    \begin{subfigure}[b]{0.49\columnwidth}
        \centering
        \includegraphics[width=\textwidth]{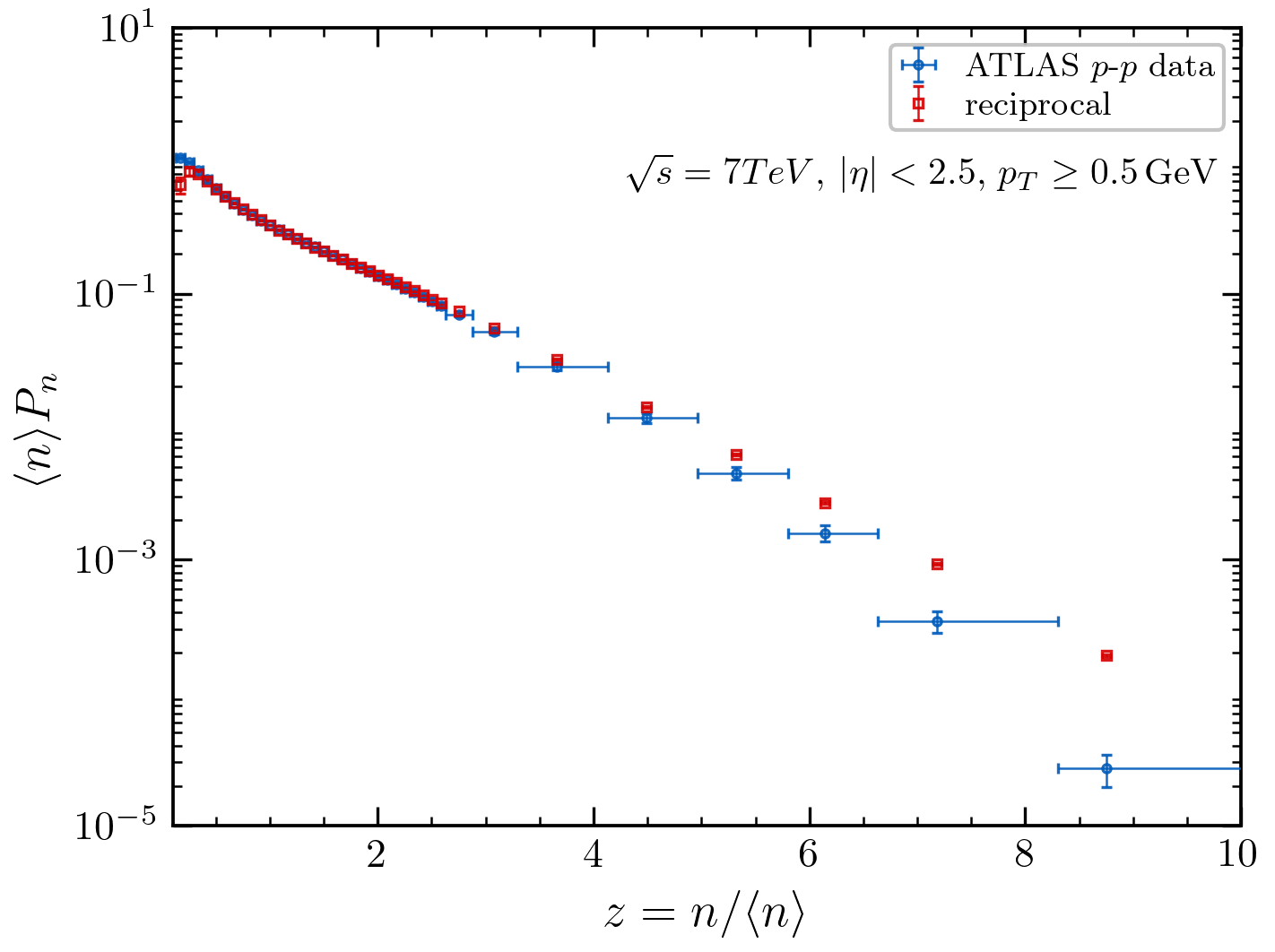}
        \caption{$\sqrt{s}=7$~TeV}
        \label{fig:psi7}
    \end{subfigure}\hfill
    \begin{subfigure}[b]{0.49\columnwidth}
        \centering
        \includegraphics[width=\textwidth]{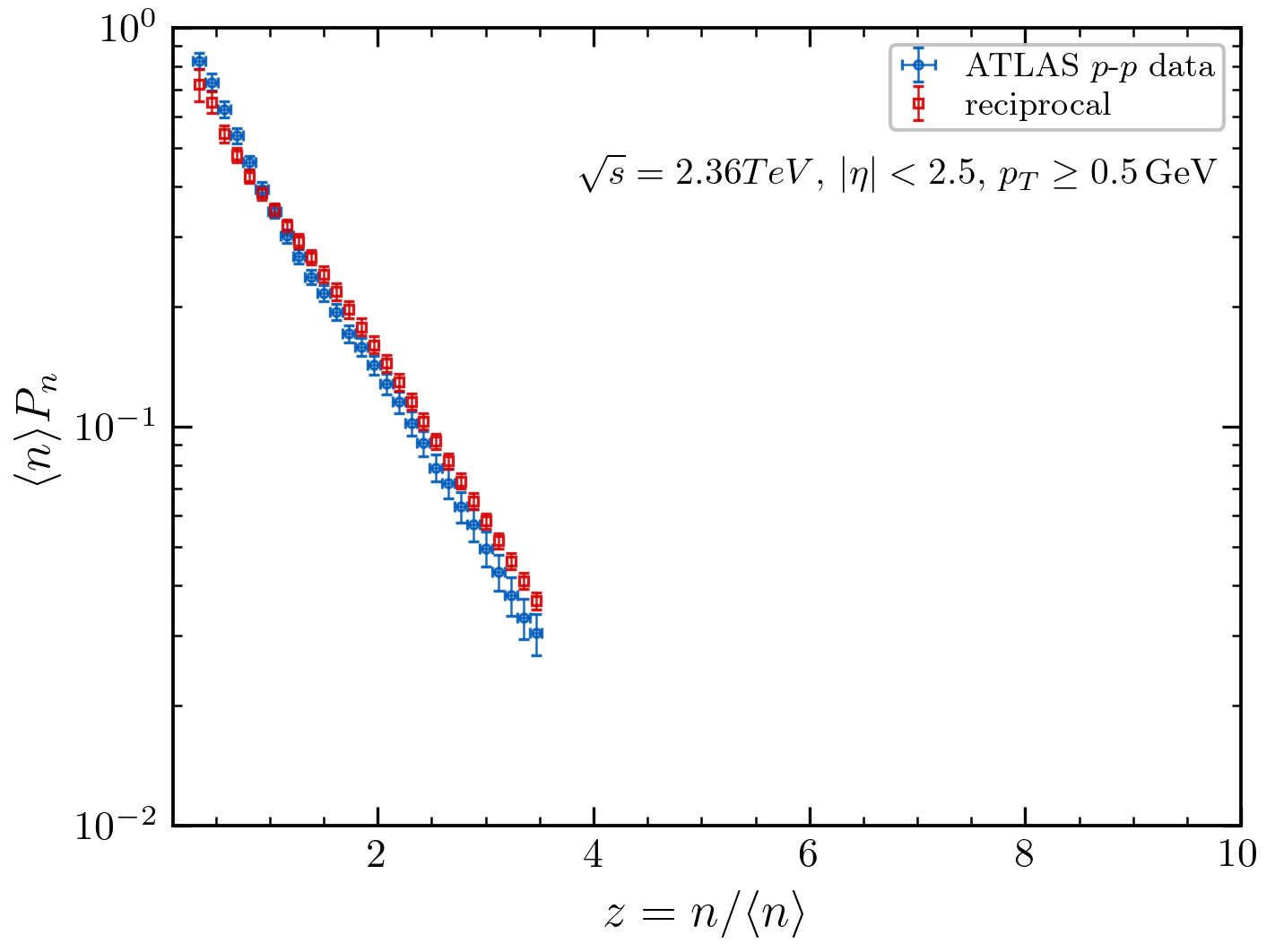}
        \caption{$\sqrt{s}=2.36$~TeV}
        \label{fig:psi236}
    \end{subfigure}
    \caption{ATLAS data plots for $p-p$ collisions at $\sqrt{s}=2.36,\,7,\,8,\,13$~TeV, $p_t>500$~MeV and $|\eta|<2.5$. The blue points represent $\langle n\rangle P_n$ as a function of $z$, and the red points represent the transformed probability $e^{1/z-z}[\langle n\rangle P_n](1/z)$ given in (\ref{nPnINV}), illustrating the reciprocal symmetry $z\leftrightarrow 1/z$ of $f_s(z)$ in the range $1/3<z<3$ at $\sqrt{s}=7,\,8,\,13$~TeV. The averages $\langle n\rangle$ are those calculated in~\cite{Kulch} and given in Table~\ref{table:average}.}
    \label{fig:nPnreciprocal}
\end{figure}

  It is instructive to reformulate the reciprocal symmetry
$z\leftrightarrow 1/z$ of $f_s(z)$ as a transformation for the KNO scaled
probability $\langle n\rangle P_n$. Using (\ref{nPnfs}) we write
\begin{eqnarray}
[\langle n\rangle P_n](z)= e^{-z}\bigl(1+f_s(z)\bigr)
\end{eqnarray}
and its reciprocal expression
\begin{eqnarray}
[\langle n\rangle P_n]\!\left(\tfrac{1}{z}\right)= e^{-1/z}\!\left(1+f_s\!\left(\tfrac{1}{z}\right)\right).
\end{eqnarray}
Provided the reciprocal symmetry holds, $f_s(z)=f_s(1/z)$, we obtain
\begin{eqnarray}
[\langle n\rangle P_n](z)= e^{1/z-z}\,[\langle n\rangle P_n]\!\left(\tfrac{1}{z}\right).
\label{nPnINV}
\end{eqnarray}
The resulting transformation of the scaled probability $\langle n\rangle P_n$
versus the original experimental data is depicted in
Fig.~\ref{fig:nPnreciprocal}. The transformed scaled probability
$e^{1/z-z}[\langle n\rangle P_n](1/z)$ reasonably reproduces the experimental
data for $[\langle n\rangle P_n](z)$. We emphasize that no fit function is
involved in Fig.~\ref{fig:nPnreciprocal}: the curves contain only the
experimental data points and their transformed counterparts using the
transformation given by (\ref{nPnINV}).

\subsection{Reciprocal symmetry in CMS data}

\begin{figure}[!htb]
    \centering
    \begin{subfigure}[b]{0.49\columnwidth}
        \centering
        \includegraphics[width=\textwidth]{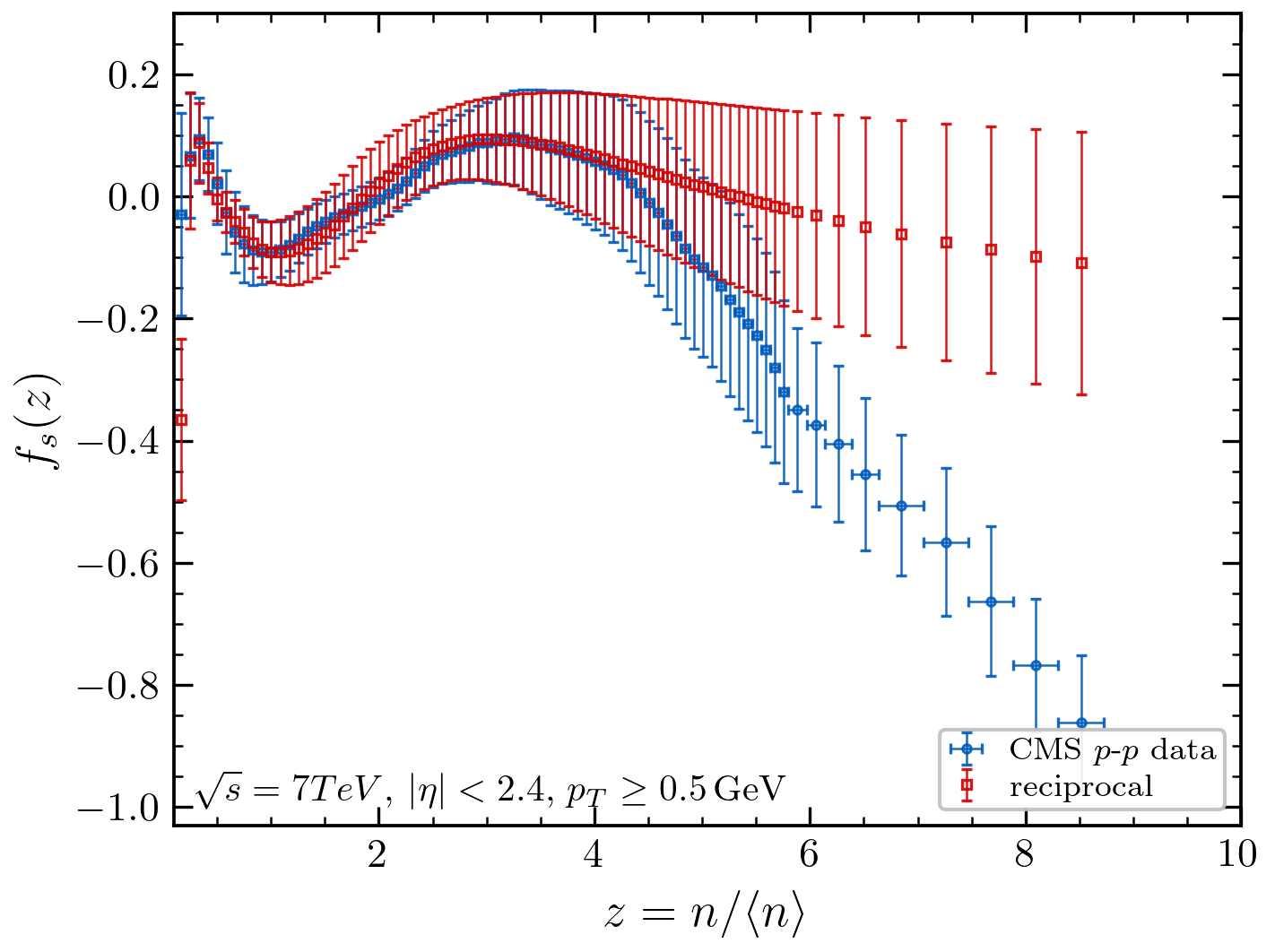}
        \caption{$f_s(z)$ at $\sqrt{s}=7$~TeV}
        \label{fig:fscms7}
    \end{subfigure}\hfill
    \begin{subfigure}[b]{0.49\columnwidth}
        \centering
        \includegraphics[width=\textwidth]{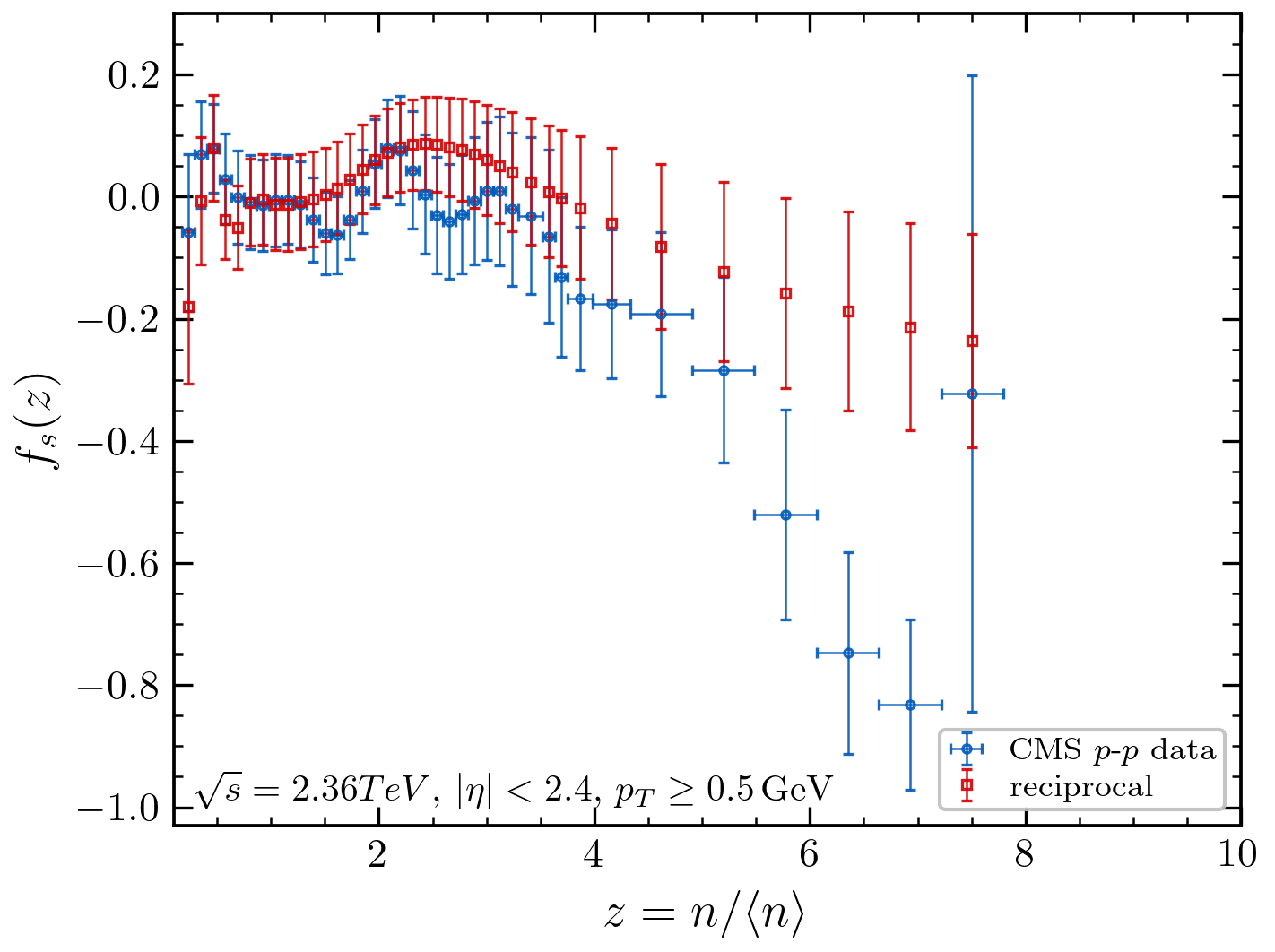}
        \caption{$f_s(z)$ at $\sqrt{s}=2.36$~TeV}
        \label{fig:fscms236}
    \end{subfigure}

    \vspace{0.4cm}

    \begin{subfigure}[b]{0.49\columnwidth}
        \centering
        \includegraphics[width=\textwidth]{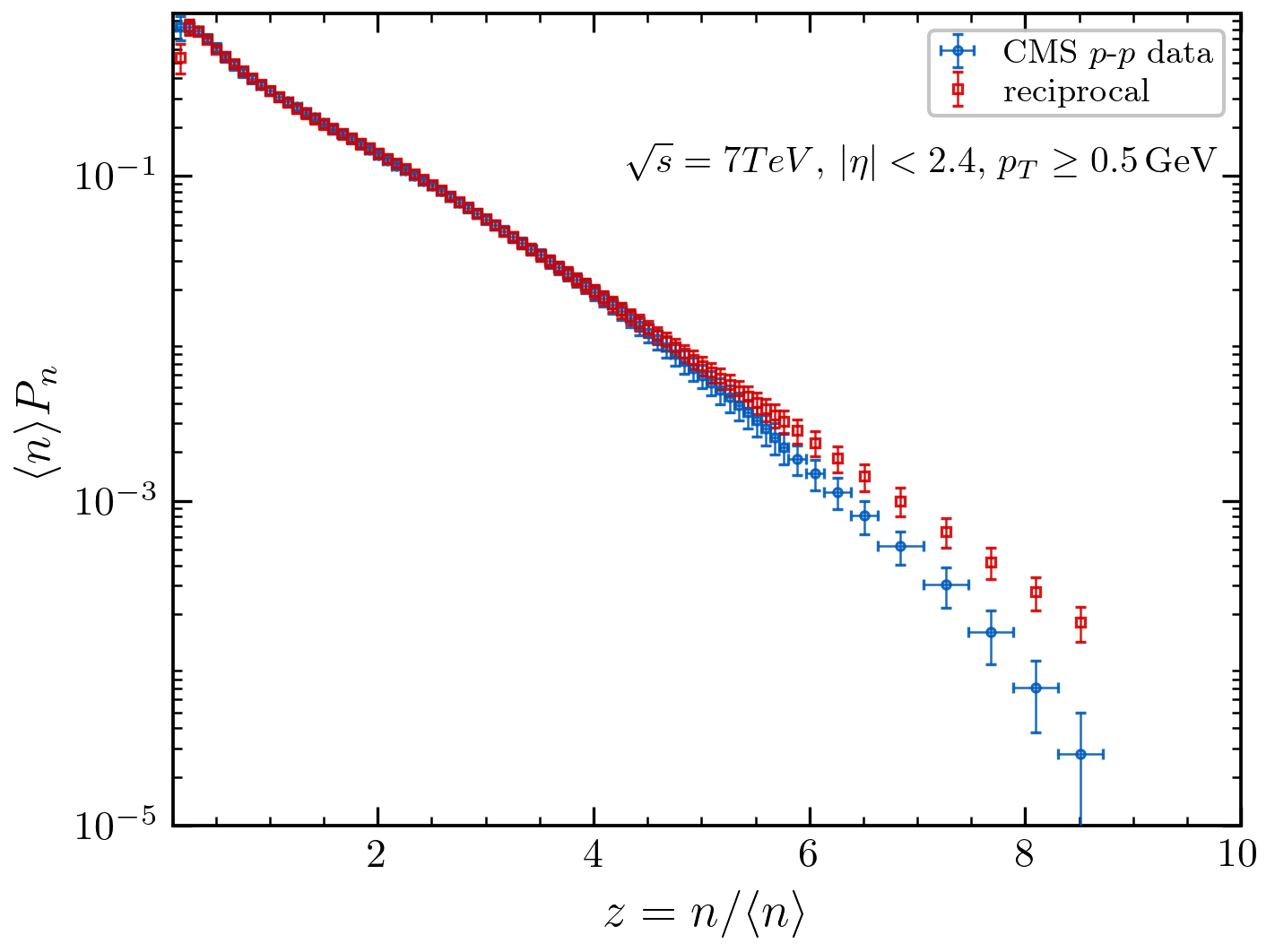}
        \caption{$\langle n\rangle P_n$ at $\sqrt{s}=7$~TeV}
        \label{fig:psicms7}
    \end{subfigure}\hfill
    \begin{subfigure}[b]{0.49\columnwidth}
        \centering
        \includegraphics[width=\textwidth]{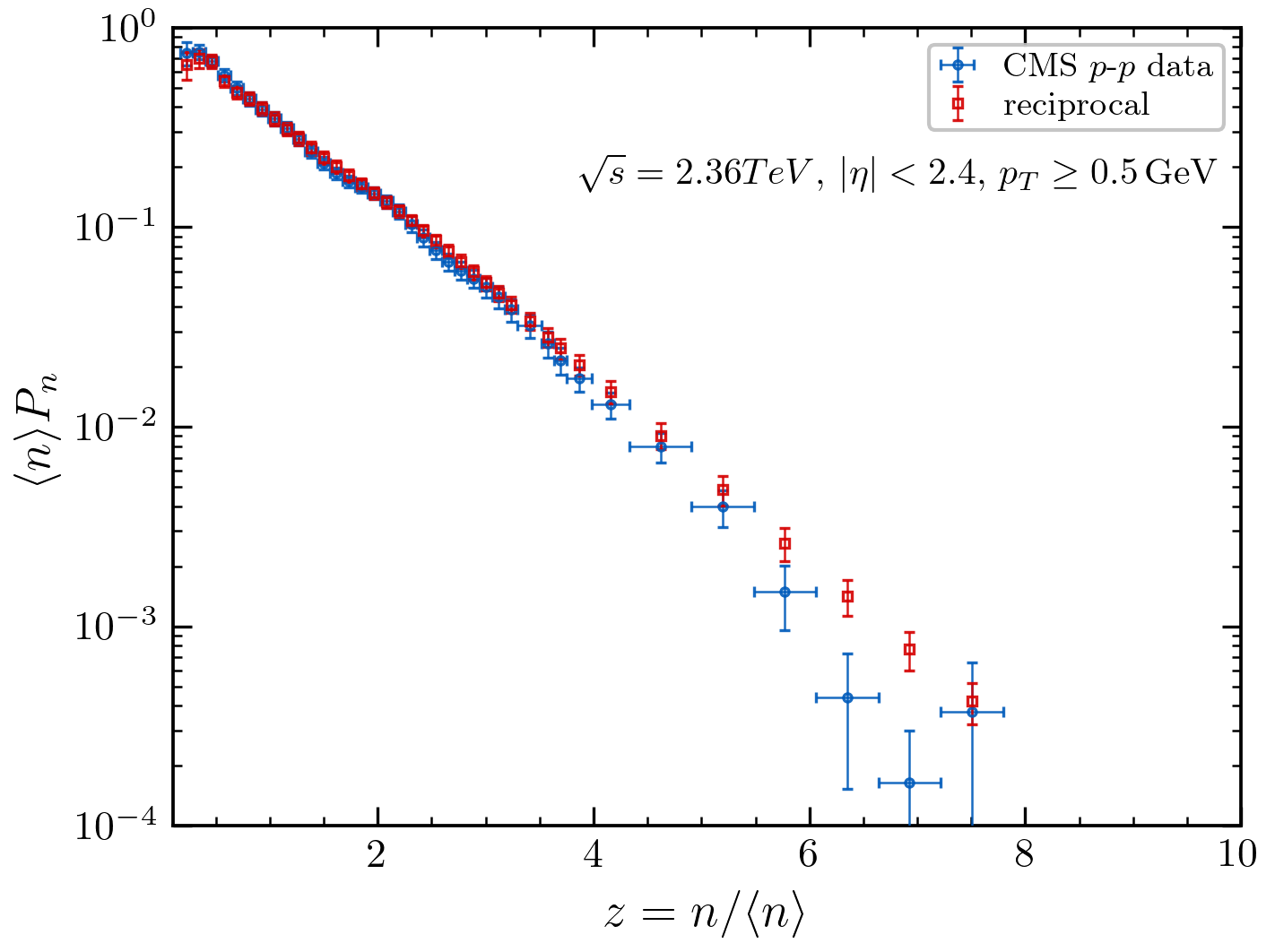}
        \caption{$\langle n\rangle P_n$ at $\sqrt{s}=2.36$~TeV}
        \label{fig:psicms236}
    \end{subfigure}

    \caption{Reciprocal symmetry in the CMS data~\cite{CMS} at $\sqrt{s}=7$~TeV and $\sqrt{s}=2.36$~TeV with $p_t>500$~MeV and $|\eta|<2.4$. Top row: $f_s(z)$ in blue and $f_s(1/z)$ in red. Bottom row: $\langle n\rangle P_n$ in blue and the transformed probability $e^{1/z-z}[\langle n\rangle P_n](1/z)$ in red. As in the ATLAS data, the reciprocal symmetry is observed at the higher energy in the range $1/3<z<3$ and is much less pronounced at $\sqrt{s}=2.36$~TeV.}
    \label{fig:nPnreciprocalCMS}
\end{figure}

   We extend the reciprocal symmetry analysis to the CMS experimental
data~\cite{CMS}. The corresponding plots for $f_s(z)$ and
$\langle n\rangle P_n$ at $\sqrt{s}=7$~TeV and $\sqrt{s}=2.36$~TeV are shown
in Fig.~\ref{fig:nPnreciprocalCMS}. 
The CMS measurements used here
correspond to $p_t>500$~MeV and $|\eta|<2.4$, and we use the corresponding
average multiplicities $\langle n\rangle$. As in the ATLAS data, the
reciprocal symmetry is observed at $\sqrt{s}=7$~TeV in the central window
$1/3<z<3$, while at $\sqrt{s}=2.36$~TeV it is much less pronounced.

\section{Local consequence of the reciprocal symmetry}
\label{sec:local}

 The reciprocal symmetry $f_s(z)=f_s(1/z)$ implies a simple, fit-independent
local relation at $z=1$. Differentiating the symmetry relation gives
\begin{eqnarray}
f'_s(z)=-\frac{1}{z^2}f'_s\!\left(\tfrac{1}{z}\right),
\end{eqnarray}
which at $z=1$ reduces to
\begin{eqnarray}
f'_s(1)=-f'_s(1)\quad\Longrightarrow\quad f'_s(1)=0.
\label{fsprime1}
\end{eqnarray}
Thus the reciprocal symmetry forces the derivative of $f_s(z)$ to vanish
at $z=1$.

On the other hand, $f_s(z)$ defined in (\ref{fs}) reads
\begin{eqnarray}
f_s(z)=-1+e^{z}\langle n\rangle P(z\langle n\rangle),
\label{fsnew}
\end{eqnarray}
where we treat $n=z\langle n\rangle$ as a continuous variable.
Differentiating with respect to $z$,
\begin{eqnarray}
f'_s(z)=\langle n\rangle e^{z}\!\left[P(z\langle n\rangle)+\langle n\rangle P'(z\langle n\rangle)\right].
\end{eqnarray}
Setting $z=1$ and using (\ref{fsprime1}) gives a direct relation between
the value and the derivative of $P(n)$ at $n=\langle n\rangle$,
\begin{eqnarray}
P'(\langle n\rangle)=-\frac{1}{\langle n\rangle}P(\langle n\rangle).
\label{PderP}
\end{eqnarray}

 
 The relation (\ref{PderP}) is a falsifiable, fit-independent prediction of
the reciprocal symmetry. To test it directly, we define the dimensionless
ratio
\begin{eqnarray}
\rho\equiv -\langle n\rangle\,\frac{P'(\langle n\rangle)}{P(\langle n\rangle)},
\label{ratio}
\end{eqnarray}
which 
equals unity if (\ref{PderP}) holds. We extract $\rho$ from the
experimental data using the local Gaussian fit
$P(n)=\frac{1}{\langle n\rangle}e^{-n/\langle n\rangle}\bigl(1+f^{\mathrm{fit}}_s(n/\langle n\rangle)\bigr)$
with $f^{\mathrm{fit}}_s$ from (\ref{fsgauss}), restricting the fit to the
ten 
data points closest to $z=1$. The resulting values of $\rho$ are
listed in Table~\ref{tab:rho}: at $\sqrt{s}=7$, $8$, and $13$~TeV, $\rho$
equals unity at the level of a few percent, providing a direct
experimental confirmation of the local relation (\ref{PderP}) and hence of
the reciprocal symmetry in the vicinity of $z=1$.

 \begin{table}[htbp]
    \centering
    \renewcommand{\arraystretch}{1.2}
    \begin{tabular}{lcccccc}
        \hline\hline
        \textbf{Energy} & $\boldsymbol{\langle n\rangle}$ & $\boldsymbol{a}$ & $\boldsymbol{b}$ & $\boldsymbol{c}$ & $\boldsymbol{\mu}$ & $\boldsymbol{\rho}$ \\
        \hline
        13 TeV & 14.66 & $\phantom{-}0.2982$ & $-0.4479$ & $0.6418$ & $-0.0331$  & $0.9776$ \\
        8 TeV  & 12.25 & $\phantom{-}0.2826$ & $-0.4009$ & $0.8518$ & $-0.0228$  & $0.9824$ \\
        7 TeV  & 11.98 & $\phantom{-}0.2858$ & $-0.3903$ & $0.7345$ & $\phantom{-}0.0017$ & $1.0011$ \\
        \hline\hline
    \end{tabular}
    \caption{Values of the ratio $\rho=-\langle n\rangle\,P'(\langle n\rangle)/P(\langle n\rangle)$ defined in (\ref{ratio}), obtained from a local Gaussian fit (\ref{fsgauss}) of $f_s(z)$ to the ten data points closest to $z=1$. The deviation of $\rho$ from unity stays below $\sim 2.3\%$ for $\sqrt{s}=7,\,8,\,13$~TeV, providing a direct experimental confirmation of the relation $P'(\langle n\rangle)=-P(\langle n\rangle)/\langle n\rangle$ in (\ref{PderP}).}
    \label{tab:rho}
\end{table}

\section{Extracting the entanglement entropy}
 \label{sec:entropy}

   The local relation (\ref{PderP}) suggests an alternative route for
extracting the entanglement entropy from the experimental data without
resorting to global fits of the multiplicity distribution. The entropy
associated with the multiplicity distribution is given by
\begin{eqnarray}
S=-\sum_{n=1}^{\infty}P(n)\,\ln P(n),
\label{vonNeumann}
\end{eqnarray}
which coincides with the von Neumann entropy of the partonic density
matrix when the latter is diagonal in the Fock basis~\cite{KharzeevLevin}.
For both the MKL and AGK models, the large-$\langle n\rangle$ expansion of
$S$ takes the form~\cite{KharzeevLevin,OuchenPrygarin1,HKStraka,H1}
\begin{eqnarray}
S\simeq \ln\langle n\rangle+1+\mathcal{O}\!\left(\frac{1}{\langle n\rangle}\right),
\label{Sasympt}
\end{eqnarray}
so that for large $\langle n\rangle$ the average multiplicity is
approximately related to $S$ by
\begin{eqnarray}
\langle n\rangle\simeq e^{S-1}.
\label{aveS}
\end{eqnarray}
We emphasize that the relation (\ref{aveS}) is model-dependent: it follows
from the leading exponential behavior of $\langle n\rangle P_n$, which is
a generic feature of the MKL and AGK models but holds only approximately
in the data. Combining (\ref{aveS}) with the local consequence
(\ref{PderP}) of the reciprocal symmetry gives
\begin{eqnarray}
P'\!\left(e^{S-1}\right)\simeq -e^{1-S}\,P\!\left(e^{S-1}\right).
\label{PS}
\end{eqnarray}
The expression (\ref{PS}) relates the local properties of the production
probability in the vicinity of $z=1$ (i.e.\ at $n\simeq\langle n\rangle$) to
the global quantity $S$. Measurements of $P_n$ have small uncertainties
near $n\simeq\langle n\rangle$, in contrast to the large experimental
errors of the distribution tail at $n\gg\langle n\rangle$ that can
significantly affect $S$ extracted from a global fit. The relation
(\ref{PS}) therefore provides a methodology for extracting $S$ from the
well-measured central region, complementary to a global-fit
determination, and allows for a more direct comparison with the
entanglement entropy $S$ calculated from parton distribution functions in
the Kharzeev--Levin framework~\cite{KharzeevLevin,Kharzeev2021}, from
BFKL-evolved unintegrated gluon
distributions~\cite{HentschinskiKutak2022,HKStraka}, or from dipole cascade
models~\cite{KutakPraszalowicz,KutakLokos}. The comparison between
partonic and hadronic entropy is sensitive to the choice of charged-vs-total
multiplicity and to the kinematic window~\cite{HKStraka,HKKT2024}, both of
which constitute important systematics for any such extraction at the LHC.
A detailed numerical extraction of $S$ along these lines is left for
future work.

\section{Conclusion}

     We have identified a reciprocal symmetry $z\leftrightarrow 1/z$ in the KNO
violating term $f_s(z)$ in the range $1/3<z<3$, observed in the ATLAS and
CMS charged-multiplicity data at $\sqrt{s}=7,\,8,\,13$~TeV. The symmetry
is well described by a Gaussian 
in $\ln z$ and becomes more pronounced
with increasing collision energy. As a direct consequence of the symmetry,
the derivative of $f_s(z)$ vanishes at $z=1$, which translates into the
local relation $P'(\langle n\rangle)=-P(\langle n\rangle)/\langle n\rangle$.
We have verified
 this relation in the data by computing the ratio
$\rho\equiv-\langle n\rangle P'(\langle n\rangle)/P(\langle n\rangle)$ and
finding $\rho=1$ at the few-percent level at all three high energies.
Combining this local relation with the large-$\langle n\rangle$
approximation $\langle n\rangle\simeq e^{S-1}$ from the MKL/AGK models
allows the entanglement entropy to be extracted from the well-measured
region $n\simeq\langle n\rangle$, avoiding the large uncertainties
associated with the
 distribution tail. The dynamical origin of the
reciprocal symmetry is left for future investigation. In particular, it
would be interesting to explore its possible connection to the conformal
structure of high-energy QCD, to the diffusion-scaling alternative to KNO
recently proposed in~\cite{MoriggiNavarraMachado}, and to the small-$x$
QCD origins of multiplicity distributions explored
in~\cite{LiuNowakZahed,Dumitru}.

\begin{acknowledgments}
We are indebted to Sergey Bondarenko for inspiring discussions. This work
is supported in part by the ``Program of HEP support -- Council of Higher
Education of Israel.''
\end{acknowledgments}


\end{document}